\documentclass[aps,prx,superscriptaddress,twocolumn]{revtex4-1}
\usepackage{bbm}
\usepackage{mathrsfs}
\usepackage{amsmath}
\usepackage{amsfonts}
\usepackage[colorlinks=true,linkcolor=blue,urlcolor=blue,citecolor=blue,anchorcolor=blue]{hyperref}
\usepackage{graphicx,epstopdf}
\usepackage{subfigure}
\usepackage{epsfig}
\usepackage{dcolumn}
\usepackage{bm}
\usepackage{color}
\usepackage{natbib}
\usepackage{amssymb}
\usepackage{xcolor}
\usepackage{braket}
\usepackage{ulem}
\usepackage{float}
\usepackage{lipsum}

\begin{document}
\title{Abelian and non-Abelian quantum spin liquids in a three-component Bose gas on optical Kagome lattices}

\author{Kaiye Shi}
\affiliation{Department of Physics and Beijing Key Laboratory of Opto-electronic Functional Materials and Micro-nano Devices, Renmin University of China, Beijing 100872, China}
\affiliation{Key Laboratory of Quantum State Construction and Manipulation (Ministry of Education), Renmin University of China, Beijing 100872,China}
\author{Wei Zhang}
\email{wzhangl@ruc.edu.cn}
\affiliation{Department of Physics and Beijing Key Laboratory of Opto-electronic Functional Materials and Micro-nano Devices, Renmin University of China, Beijing 100872, China}
\affiliation{Key Laboratory of Quantum State Construction and Manipulation (Ministry of Education), Renmin University of China, Beijing 100872,China}
\address{Beijing Academy of Quantum Information Sciences, Beijing 100193, China}
\author{Zheng-Xin Liu}
\email{liuzxphys@ruc.edu.cn}	
\affiliation{Department of Physics and Beijing Key Laboratory of Opto-electronic Functional Materials and Micro-nano Devices, Renmin University of China, Beijing 100872, China}
\affiliation{Key Laboratory of Quantum State Construction and Manipulation (Ministry of Education), Renmin University of China, Beijing 100872,China}

\begin{abstract}
Realization of non-Abelian anyons in topological phases is a crucial step toward topological quantum computation. We propose a scheme to realize a non-Abelian quantum spin liquid (QSL) phase in a three-component Bose gas with contact interaction on optical Kagome lattices. In the strong coupling regime, the system is described by an effective spin-1 model with two- and three-body interactions between neighboring spins. By mapping out the phase diagram via variational Monte Carlo method, we find a non-Abelian chiral spin liquid phase in which the Ising-type anyons obey non-Abelian braiding statistics. The gapless chiral edge states can be detected by measuring the spin-spin correlation from atomic population. Furthermore, an interesting $Z_2$ QSL phase is observed exhibiting both topological order and lattice symmetry breaking order. Our scheme can be implemented in cold quantum gases of bosonic atoms.
\end{abstract}

\maketitle

\section{Introduction}
\label{sec:intro}
The investigation of topological quantum matter~\cite{Thouless1982,Wen1990T} has significantly broadened the scope of quantum phases beyond the symmetry breaking paradigm. As one important type of strongly correlated topological phases, the quantum spin liquid (QSL)~\cite{Zhou2017} has attracted a lot of attention since it was proposed~\cite{Anderson1987,Wen2017}. A gapped QSL with non-trivial intrinsic topological order~\cite{Wen1989, Wen1990} is a long-range entangled state~\cite{Chen2010} with emergent gauge fields, which is characterized by topological degeneracy of ground states on a torus and fractionalized bulk excitations referred as anyons. If the anyons obey non-Abelian braiding statistics~\cite{Greiter2009, Kitaev2006} like Majorana zero modes in topological superconductors, they can be utilized in fault-tolerant quantum computation~\cite{Nayak2008}. The QSLs are likely to be realized in frustrated antiferromagnets, owing to the strong quantum fluctuations therein~\cite{Savary2017}. Thus, as one of the most frustrated structures in two dimensions, the Kagome lattice is considered to be a promising candidate for both spin-1/2 and spin-1 systems~\cite{Han2012, Ran2007, Liu2018, Cheng2011,Fu2015}. Although some indications of QSLs have been reported in solid state materials~\cite{Han2012,Cheng2011,Fu2015}, a direct experimental detection without any ambiguity is difficult to obtain. 

 On the other hand, due to the remarkable controllability and flexibility, ultracold quantum gases of neutral atoms provide a versatile platform for quantum simulation of these exotic phases~\cite{Bloch2008, Eckardt2017}. In the past decade, rapid progresses have been made in realizing topological bands of different types using ultracold atoms without interaction~\cite{Cooper2019,Jotzu2014,Song2018}. Meanwhile, since the cold atom platform can flexibly adjust the trapping configuration and interatomic interactions~\cite{Bloch2008}, many important lattice models have been realized, including the Hubbard model and the Heisenberg antiferromagnets~\cite{Greiner-02,Mazurenko2017}. These progresses make the realization of strongly interacting topological phases accessible~\cite{Goldman2016}.

Only until lately, a simulation of QSL with $Z_2$ topological order has been reported in an array of Rydberg atoms~\cite{Semeghini2021}.  As the first attempt to realize topological orders in an intrinsically bosonic system, this work triggers a wave of follow-up theoretical studies on the QSL~\cite{Giudici2022,Tarabunga2022}. Up to date, much effort has been devoted to Abelian QSLs, while the realization of gapped non-Abelian QSLs with anyons obeying non-Abelian braiding statistics is still a challenging task yet to be accomplished. Recently, two proposals to implement Floquet non-Abelian QSLs in cold atoms under periodically driving have been raised~\cite{Kalinowski2023,Sun2023}. However, Floquet engineering usually brings inevitable heating effect, which may hamper the realization of the ideal Floquet Hamiltonian in realistic experimental conditions~\cite{Rudner2020}. Therefore, a non-Abelian QSL as an equilibrium state is of great importance in aspects of both fundamental physics and potential applications.

Here, we propose to realize a non-Abelian QSL in a three-component Bose gas of ultracold atoms on an optical Kagome lattice. Using the technique of Raman coupling, we can tune the phase of hopping integrals and implement non-zero fluxes. In the weak hopping limit, an effective $S=1$ spinor model is derived when each site is filled by one particle only. We then use the variational Monte Carlo (VMC) method to map out the phase diagram and find several $Z_2$ spin liquid ($Z_2$-SL) with coexisting symmetry breaking orders, and most interestingly a non-Abelian chiral QSL phase, which can be experimentally identified from spin-spin correlations. Our proposal can be implemented with existing techniques and provide a practical guideline for experimental implementation of non-Abelian anyons in equilibrium systems. The remainder of the manuscript is organized as follows. In Sec.~\ref{sec:BHModel}, we introduce the proposed experimental scheme of spin-1 Bose gases in optical Kagome lattice and derive the effective Hamiltonian in the weak hopping limit. In Sec.~\ref{sec:phasediagram}, we use a variational quantum Monte Carlo method to solve for the ground state of the effective spin model and map out the phase diagram. Considering the constraint of physical parameters to validate the effective spin model, we analyze the existence and characteristics of various QSL phases for the original lattice Hamiltonian in Sec.~\ref{sec:QSL}. Finally, we summarize in Sec.~\ref{sec:summary}
\begin{figure}[tbp]
	\centering
	\includegraphics[width=1\linewidth]{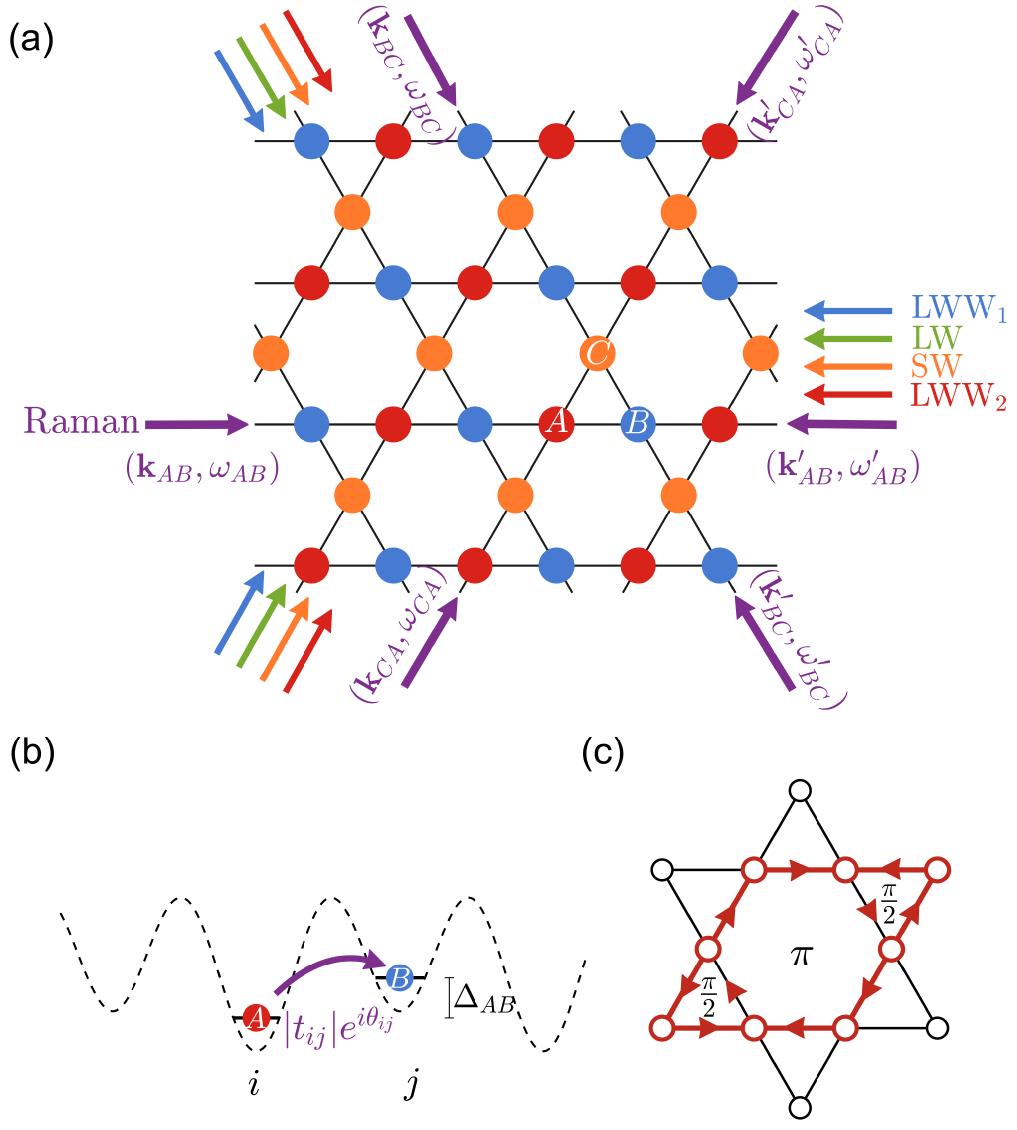}
	\caption{(a) A Kagome lattice is constructed by staggered triangular lattices of short- (SW) and long-wavelength (LW) lasers. Two additional long-wavelength lattices of weak intensity LWW$_1$ and LWW$_2$ are introduced to lift the degeneracy of the three sublattices. (b) Complex hopping coefficients are realized by counter-propagating Raman lasers $\mathbf{k}$ and $\mathbf{k}'$. (c) When the Raman beams are tuned to give a hopping with phase $e^{i\pi/6}$ along all arrows, a pattern with the same flux $\pi/2$ in triangles ($\bigtriangleup$) and inverse triangles ($\bigtriangledown$) can be realized.}
	\label{fig1} 
\end{figure}
%

\section{The Bose-Hubbard model with complex hopping for spinor bosons}
\label{sec:BHModel}

\subsection{Experimental setup}
\label{sec:BHModel:set}

We consider a three-component Bose-Einstein condensate~\cite{Zhao2015} in the three hyperfine states of the $S=1$ manifold adiabatically loaded on an optical Kagome lattice, which is composed by two staggered triangle lattices using linearly polarized short-wavelength (SW) 532nm and long-wavelength (LW) 1064nm lasers~\cite{Jo2012}, as illustrated in Fig.~\ref{fig1}(a). The second quantized Hamiltonian takes the form (with $\hbar=1$)
\begin{eqnarray}
	\label{eq1}
	H&=&\sum_{\sigma}\int d\bm{r}\psi_{\sigma}^{\dagger}(\bm{r})
	\left [-\frac{\bigtriangledown^2}{2m}+V_L(\bm{r})+V_T(\bm{r})\right] \psi_{\sigma}(\bm{r}) \nonumber \\
	+ &&\frac{1}{2}\sum_{\sigma_{\delta},\sigma_{\xi},\sigma_{\gamma},\sigma_{\eta}}\int V_{\sigma_{\delta}\sigma_{\xi}\sigma_{\gamma}\sigma_{\eta}}(\bm{r}_1,\bm{r}_2)\psi_{\sigma_{\delta}}^{\dagger}(\bm{r}_1) \psi_{\sigma_{\xi}}^{\dagger}(\bm{r}_2)\nonumber \\ &&\psi_{\sigma_{\gamma}}(\bm{r}_2)\psi_{\sigma_{\eta}}(\bm{r}_1)d\bm{r}_1d\bm{r}_2.
\end{eqnarray}
Here, $\psi_{\sigma}(\bm{r})$ is the atomic field annihilation operator for atoms in the pseudo spin state $\ket{S=1,S^z=\sigma}$ with $\sigma=\pm 1,0$, $V_L(\bm{r})$ is the Kagome lattice potential, $V_T(\bm{r})$ is a global trapping potential, $V_{\sigma_{\delta}\sigma_{\xi}\sigma_{\gamma}\sigma_{\eta}}(\bm{r}_1,\bm{r}_2)={_{1}\bra{\sigma_{\delta}}}{_{2}\bra{\sigma_{\xi}}}\hat{V}(\bm{r}_1,\bm{r}_2)\ket{\sigma_{\gamma}}_2\ket{\sigma_{\eta}}_1$ describes the effective two-body interaction between atoms, with $\hat{V}(\bm{r}_1,\bm{r}_2)$ in the form of a contact potential $	\hat{V}(\bm{r}_1,\bm{r}_2)=\delta(\bm{r}_1-\bm{r}_2) \big(g_0\hat{P}_0+g_2\hat{P}_2 \big)$ ~\cite{Imambekov2003}, where $\hat{P}_{\mathcal{S}}$ is the projection operator for the pair of atoms with total spin $\mathcal{S}$, and the interaction strength $g_{\mathcal{S}}=4\pi \hbar^2 a_{\mathcal{S}}/m$ with $a_{\mathcal{S}}$ the $s$-wave scattering length for the spin-$\mathcal{S}$ channel. 

When the temperature and interaction strength are much smaller than the band gap, the effect of high-energy bands can be ignored and the atoms are confined to the lowest Bloch band, leading to a generalized Hubbard model with multiple components 
\begin{eqnarray}
	\label{eqn:BHmodel}
	\mathcal{H}_{\mathrm{latt}}&=&\frac{U_0}{2}\sum_i\hat{n}_i(\hat{n}_i-1)-\sum_i\mu\hat{n}_i+\frac{U_2}{2}\sum_i(\bm{\mathrm{S}}^2_i-2\hat{n}_i) \nonumber \\
	&&-t_{\mathrm{kin}}\sum_{\langle i,j\rangle,\sigma}(a_{\sigma j}^{\dagger}a_{\sigma i}+a^{\dagger}_{\sigma i}a_{\sigma j}).
\end{eqnarray}
Here, $a_{\sigma i}$ is the annihilation operator for state in the lowest band localized on site $i$ and with spin component $\sigma=\{\pm 1,0\}$, $\hat{n}_i=\sum_{\sigma}a_{\sigma i}^{\dagger}a_{\sigma i}$ is the total number of particles on site $i$, and $\mu$ is the chemical potential. In the interaction terms, $U_0=\frac{g_0+2g_2}{3}\int d\bm{r}|w(\bm{r})|^4$ represents the Hubbard repulsion, $U_2=\frac{g_2-g_0}{3}\int d\bm{r}|w(\bm{r})|^4$ is derived from the difference in the scattering lengths of $\mathcal{S}=0$ and $\mathcal{S}=2$ channels, $w(\bm{r})$ is the Wannier basis of the lowest band, and $\bm{\mathrm{S}}_i=\sum_{\sigma\sigma'}a^{\dagger}_{i\sigma}\bm{\mathrm{L}}_{\sigma\sigma'}a_{i\sigma'}$ represents the total spin on site $i$ with $\bm{\mathrm{L}}$ the $3\times 3$ matrix representations of spin operators. The last term $t_{\mathrm{kin}}$ represents the bare kinetic tunneling between nearest neighbors $\langle i,j \rangle$.

To introduce non-zero fluxes through the triangles of Kagome lattices [Fig.~\ref{fig1}(c)], the hopping coefficients need to carry a phase. This can be induced by either Raman coupling~\cite{Aidelsburger2011, Jaksch2003} or periodically shaking the lattice~\cite{Jotzu2014}.  Here, we choose the Raman method since it is relatively easy to implement and adjust. Specifically, two additional triangle lattices are imposed by applying linearly polarized long-wavelength 1064nm lasers (LWW$_1$ and LWW$_2$) with weak intensity, such that the three sublattices labeled by A, B and C feel different potential depths. The staggered potential with energy offset $\Delta_{AB}$ along the AB direction is illustrated in Fig.~\ref{fig1}(b) as an example, which suppresses the kinetic tunneling between nearest-neighboring sites. Then, three pairs of Raman lasers are used to induce a nearest-neighbor tunneling $t_{ij} = |t_{ij} | e^{i\theta_{ij}}$ [Fig.~\ref{fig1}(b)], where the frequency difference of the two Raman beams is approximately equal to the energy offset of two corresponding sublattices, such as $\omega_{AB}-\omega'_{AB}\simeq \Delta_{AB}<<U_0$. Thus, the Raman tunneling term $\mathcal{H}_t=-\sum_{\langle i,j\rangle,\sigma}(t_{ij}a_{\sigma j}^{\dagger}a_{\sigma i}+t_{ji}a^{\dagger}_{\sigma i}a_{\sigma j})$ will replace the kinetic tunneling term in the Eq.~(\ref{eqn:BHmodel}), making the final Hamiltonian expressed as
\begin{eqnarray}
	\label{eqn:spinmodel}
	\mathcal{H}&=&\frac{U_0}{2}\sum_i\hat{n}_i(\hat{n}_i-1)-\sum_i\mu\hat{n}_i+\frac{U_2}{2}\sum_i(\bm{\mathrm{S}}^2_i-2\hat{n}_i) \nonumber \\
	&&-\sum_{\langle i,j\rangle,\sigma}(t_{ij}a_{\sigma j}^{\dagger}a_{\sigma i}+t_{ji}a^{\dagger}_{\sigma i}a_{\sigma j}).
\end{eqnarray}
The amplitude $|t_{ij}|$ and the phase $\theta_{ij}$ of the hopping coefficient $t_{ij}$ can be adjusted by tuning the intensity, the wave vector and the initial phase of the Raman lasers, as explained in Appendix~\ref{app:Raman}. By tuning $|t_{ij}|=t$ and $|\theta_{ij}|=\theta={\pi}/{6}$, we can realize non-zero fluxes through the lattice. As shown in Fig.~\ref{fig1}(c), when hopping around a closed triangular loop ($ijk$) counter-clockerwisely, the boson obtains an Aharonov-Bohm phase $\phi_{ijk}={\pi}/{2}$.

\subsection{Effective spin-model}
\label{sec:spin1}

By raising up the lattice potential adiabatically, we can reach the strong coupling limit with negligible hopping  $t \ll U_{0,2}$, where the number of particles at each site is conserved and the single-site states can be expressed as $\ket{n_i,S_i,S_i^z}$. With a properly chosen chemical potential $\mu$, we can achieve singly occupied states with $n_i=1$ for all $i$, which are the eigenstates of $\mathcal{H}_0$ with eigenenergy $E_0$. In this case, the notation $\ket{n_i,S_i,S_i^z}=\ket{1,1,\sigma_i}$ can be further simplified to $\ket{\sigma_i}$. 

In the limiting case of $t=0$, the Hamiltonian Eq.~(\ref{eqn:spinmodel}) reads
\begin{equation}
	\mathcal{H}=\mathcal{H}_0 = \frac{U_0}{2}\sum_i\hat{n}_i(\hat{n}_i-1)-\sum_i\mu\hat{n}_i+\frac{U_2}{2}\sum_i(\vec{S}^2_i-2\hat{n}_i)
\end{equation}
with a ground state energy $E_0$. Then we consider a small but nonetheless finite $t$ and treat the hopping term perturbatively, we obtain an effective spin-Hamiltonian $\mathcal{H}_{\mathrm{eff}}=\sum_n\mathcal{H}^{(n)}_{\mathrm{eff}}$, where $\mathcal{H}^{(n)}_{\mathrm{eff}}$ represents the $n$-order term. For the Hilbert space with single occupancy, since hopping term $\mathcal{H}_t$ changes the site distribution of particle number, the first-order term is zero, i.e., $\bra{\{\sigma'_i\}}\mathcal{H}^{(1)}_{\mathrm{eff}}\ket{\{\sigma_i\}}=\bra{\{\sigma'_i\}}\mathcal{H}_t\ket{\{\sigma_i\}}=0$, where $\ket{\{\sigma_i\}}=\ket{\sigma_1,\sigma_2,...,\sigma_N}$ is the ground state with single occupancy. Thus, in order to respect the breaking of time-reversal symmetry, we go to the third-order of perturbation to obtain
\begin{equation}
	\bra{\{\sigma'_i\}}	\mathcal{H}_{\mathrm{eff}}\ket{\{\sigma_i\}}=\sum^3_{n=1}\bra{\{\sigma'_i\}} \left(\mathcal{H}_t\frac{1-P_0}{E_0-\mathcal{H}_0} \right)^{n-1}\mathcal{H}_t\ket{\{\sigma_i\}}.
\end{equation}
Here, $\mathcal{H}_t=-\sum_{\langle i,j\rangle,\sigma}(t_{ij}a_{\sigma i}^{\dagger}a_{\sigma j}+t_{ji}a^{\dagger}_{\sigma j}a_{\sigma i})$, and $P_0=\sum_{\{\sigma_i\}}\ket{\{\sigma_i\}}\bra{\{\sigma_i\}}$ is the projection operator for the subspace of ground states. Considering the single occupancy constraint $\sum_{\sigma}a_{i\sigma}^{\dagger}a_{i\sigma}=1$, we get an SO(3) symmetric effective spin-1 model (up to a constant)
\begin{eqnarray}
	\label{eqn:spinmodel2}
		{\cal H}_{\mathrm{eff}} &=& \sum_{\langle i,j\rangle} \left[-J_1\bm{\mathrm{S}}_i \cdot \bm{\mathrm{S}}_j-J_2(\bm{\mathrm{S}}_i \cdot \bm{\mathrm{S}}_j)^2 \right] +
		\nonumber 
		\\&&
		\sum_{\bigtriangleup,\bigtriangledown}\sin{\phi}_{ijk}[J_r R_{ijk}-J_{\chi}(\bm{\mathrm{S}}_i\times\bm{\mathrm{S}}_j)\cdot\bm{\mathrm{S}}_k],
\end{eqnarray}
where $R_{ijk}=(\bm{\mathrm{S}}_i\times\bm{\mathrm{S}}_j)\cdot\bm{\mathrm{S}}_k \left[{1\over2}(\bm{\mathrm{S}}_i+\bm{\mathrm{S}}_j +\bm{\mathrm{S}}_k)^2-2\right]$ is the SU$(3)$ symmetric ring-exchange interaction, and the coefficients $J_1$, $J_2$, $J_r$ and $J_\chi$ are determined by hopping $t$ and interactions $U_{0,2}$, namely $J_1=\frac{2t^2}{U_+}, J_2 = \frac{2t^2}{3U_+}+\frac{4t^2}{3U_-},   J_r =  \frac{4t^3}{(U_+)^2}+\frac{2t^3}{3(U_-)^2} + \frac{4t^3}{3 U_+ U_-} -\frac{4t^3U^2_2}{(U_+ U_-)^2},  J_{\chi} = \frac{2t^3U^2_2 }{(U_+U_-)^2} + \frac{8t^3U_2}{(U_+)^2U_-}
$  with $U_\pm=(U_0-{U_2\over2})\pm{3U_2\over2}$. Further details about the experimental proposal and the derivation of effective Hamiltonians can be found in the Supplemental Material~\cite{supplementary}. 
The sign of $U_2$ can be changed by choosing proper atom species. For example, $U_2$ is positive for Na (hence  $|J_1|<|J_2|$) and negative for Rb (hence $|J_1|>|J_2|$).  In the following, we will investigate the ground states of ${\cal H}_{\mathrm{eff}}$ to approach the original Hubbard model (\ref{eqn:spinmodel}).

\section{Ground states of the effective spin model}
\label{sec:phasediagram}

In this section, we use the VMC method to calculate the ground state of the effective spin model Eq.~(\ref{eqn:spinmodel2}) and compare it with the results of exact diagonalization. It is worth noting that the phase diagram of the spin model is not equal to the phase diagram of the real system, and the experimental relevance of effective spin model is only restricted in some parameter regimes as discussed in the Sec.~\ref{sec:QSL}. However, the phase diagram of the spin model is an interesting problem by itself and can have applications in other physical realizations.

\subsection{VMC Method}
\label{sec:spinmodel-VMC}
The overview of using VMC method to solve the spin model Eq.~(\ref{eqn:spinmodel2}) is as follows. First, the Hamiltonian ${\cal H}_{\mathrm{eff}}$ is rewritten in terms of three fermionic slave particles $C_i=(c_{1i},c_{0i},c_{-1i})^{\rm T}$ satisfying the local particle-number constraint $c^{\dagger}_{1i}c_{1i}+c^{\dagger}_{0i}c_{0i}+c^{\dagger}_{-1i}c_{-1i}=1$~\cite{Liu2010}. The two-body and three-body spin interactions can be written as
\begin{eqnarray}
		\label{M:mean}
		\mathbf{S}_i\cdot \mathbf{S}_j &=& - (\hat{\chi}^{\dagger}_{ij}\hat{\chi}_{ij}+\hat{\Delta}^{\dagger}_{ij}\hat{\Delta}_{ij} ), \nonumber 
		\\
		(\mathbf{S}_i\cdot\mathbf{S}_j)^2 &=& \hat{\Delta}^{\dagger}_{ij}\hat{\Delta}_{ij}, \nonumber
		\\
		(\mathbf{S}_i\times\mathbf{S}_j)\cdot \mathbf{S}_k &=& i ( {\hat{\chi}_{ij}\hat{\Delta}^{\dagger}_{jk}\hat{\Delta}_{ki}+\hat{\Delta}_{ij}\hat{\chi}_{jk}\hat{\Delta}^{\dagger}_{ki}} + 
		\nonumber \\
		&& {\hat{\Delta}^{\dagger}_{ij} \hat{\Delta}_{jk}\hat{\chi}_{ki}-\hat{\chi}_{ij}\hat{\chi}^{\dagger}_{jk}\hat{\chi}_{ki} } )-\mathrm{H.c.},
\end{eqnarray}
where $\hat{\chi}_{ij}=\sum_{\sigma}c^{\dagger}_{\sigma i}c_{\sigma j}$ is the spinon hopping operator, and $\hat{\Delta}_{ij}=c_{1i}c_{-1j}+c_{-1i}c_{1j}-c_{0i}c_{0j}$ is the spinon singlet pairing operator. The $J_r$ term in Eq.~(\ref{eqn:spinmodel2}) preserving the SU(3) symmetry depicts a ring-exchange interaction, which is invariant under any interchange of spin components. Thus, we write the $J_r$ term as
\begin{eqnarray}
	\label{hr}
	\mathcal{H}_{r}=J_r \sum_{\bigtriangleup,\bigtriangledown}i\sin{\phi_{ijk}}\left(\hat{\chi}_{ij}\hat{\chi}_{jk}\hat{\chi}_{ki}  -\mathrm{H.c.}\right).
\end{eqnarray}
To start the variational process, it is further reduced to a noninteracting form at the mean-field level, $\mathcal{H}^{\rm qsl}_{\mathrm{MF}}  =  \sum_{\langle i,j \rangle}  \left[(\chi_{ij} C^{\dagger}_{ i}C_{ j}  +   \Delta_{ij} C^{\dagger}_{ i}\bar{C}_{ j})   + \mathrm{H.c.}\right]     + \sum_i  \lambda C^{\dagger}_{ i}C_{ i}.$ Here, $\bar{C}_i=(c^{\dagger}_{-1i},-c^{\dagger}_{0i},c^{\dagger}_{1i})^{\rm T}$, $\chi_{ij}$ is the hopping parameter, $\Delta_{ij}$ is the pairing parameter with a phase factor $\varphi_{ij}$, and the Lagrangian multiplier $\lambda$ is introduced as the effective fermionic chemical potential. The ground state of the Hamiltonian $\mathcal{H}^{\rm qsl}_{\mathrm{MF}} $ is the Bardeen-Cooper-Schrieffer (BCS) wave function with $(p_x+ip_y)$ pairing symmetry. In order to satisfy the local particle-number constraint, we perform Gutzwiller projection to the BCS wave function. A uniform solution with $\chi_{ij}=\chi$ and $\Delta_{ij}=\Delta e^{i\varphi_{ij}}$ suggests a possible solution of quantum spin liquid state, which can be either Abelian or non-Abelian~\cite{Liu2018}.

To determine the true ground state of the system, it is necessary to quantitatively analyze the competition with classical symmetry breaking orders, such as ferromagnetic (FM) or spin-nematic phases~\cite{Liu2018,Imambekov2003,Liu2015}. And one must select a reasonable set of variational parameters based on physical arguments to make the comparison trustworthy. Here, we get insights from the results of exact diagonalization for a small-size system with four unit cells ($L_x\times L_y=2\times 2$)~\cite{supplementary}, and consider the following mean-field Hamiltonian with variational parameters
\begin{eqnarray}
\label{eqn:mean-field}	
\mathcal{H}_{\mathrm{MF}}=\mathcal H_{\rm MF}^{\rm qsl} + D\sum_i (S^z_i)^2 + M \sum_i S^z_i,
\end{eqnarray}
where $(S^z_i)^2=c^{\dagger}_{1i}c_{1i}+c^{\dagger}_{-1i}c_{-1i}$, and $M$ and $D$ are the background fields inducing the FM and nematic orders, respectively. Besides, there is an additional lattice symmetry breaking parameter $T=\chi_{ij}/\chi_{kl}=\Delta_{ij}/\Delta_{kl}$ characterizing the ratio of the bond strengths between the triangle (with $i,j\in \bigtriangleup$) and the inverse triangle ($k,l\in \bigtriangledown$). With all that, we denote the mean-field ground state as $\ket{{\rm MF}(\alpha)}$ and the projected state as $\ket{\Phi(\alpha)}=P_G\ket{\rm MF(\alpha)}$, where $P_G$ stands for the Gutzwiller projection and $\alpha=(\chi,\Delta,\lambda,D,M,T)$ are the variational parameters. The optimal solution of $\alpha$ is obtained by minimizing the energy functional for the projected states $E(\alpha)= \bra{\Phi(\alpha)} {\cal H}_{\mathrm{eff}}\ket{\Phi(\alpha)}/\langle \Phi(\alpha)|  \Phi(\alpha)\rangle$. 
 
\subsection{Phase diagram of the effective spin-model}
\begin{figure}[tbp]
		\centering
		\includegraphics[width=1\linewidth]{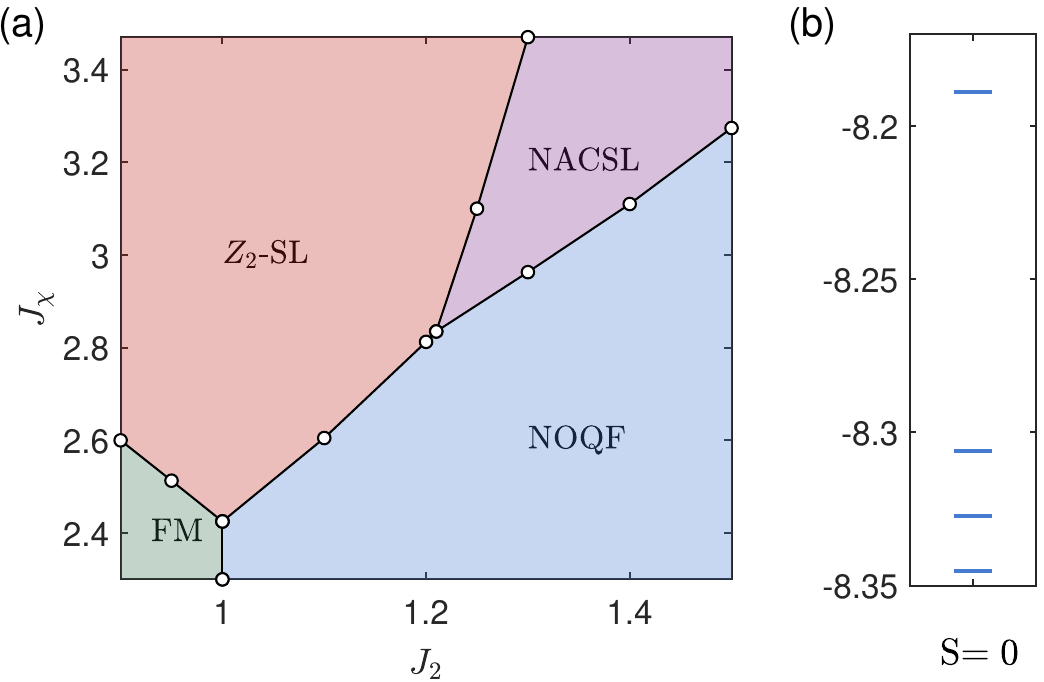}
		\caption{(a) The phase diagram of the spin-model Eq.~(\ref{eqn:spinmodel2}) with fixed $J_r=0.5$, $J_1=1$ and uniform $\phi_{ijk}=\pi/2$. (b) The lowest four energies of $S=0$ obtained by exact diagonalization of a system with $2\times2$ unit cells. The parameters are $(J_1,J_2,J_{\chi},J_r) = (1,1.9,2.72,1.6)$.}
		\label{fig2} 
\end{figure}

For the Kagome pattern of aligned fluxes as given in Fig.~\ref{fig1}(c), the effective spin-model Eq.~(\ref{eqn:spinmodel2}) preserves SO(3) spin rotation symmetry and inversion symmetry of the lattice. The VMC results of the spin model without the constraint of the Hubbard-model Eq.~(\ref{eqn:spinmodel}) parameters is shown in Fig.~\ref{fig2}(a), which presents several possible phases, including nematic order with quantum fluctuations (NOQF), ferromagnetic order (FM), non-Abelian chiral spin liquid (NACSL) and “trimerized” Abelian $Z_2$-SL. When the three-body interaction terms are small, we obtain a nematic ordered phase with quantum fluctuations (NOQF) for $J_1 < J_2$. When $J_1 > J_2$, we obtain a ferromagnetic (FM) phase with completely polarized spins. When the $J_{\chi}$ term is large enough, a $Z_2$ spin liquid ($Z_2$-SL) with large $T$ is found. This finding is rather surprising since the  $J_{\chi}$ term is considered to favor the non-Abelian chiral spin liquid (NACSL)~\cite{Liu2018}. The most intriguing NACSL phase can be achieved with $J_2 > J_ 1$~\cite{supplementary}, which corresponds to the region of $0 < U_2 < 0.5U_0$~\cite{Rizzi2005,Imambekov2003} in the original Hubbard model Eq.~(\ref{eqn:spinmodel}). The corresponding VMC results and data analysis for these phases can be found in Appendix~\ref{app:SD}. 

\begin{table*}[t]
\caption{\label{ED} 
 Different quantities obtained by exact diagonalization for a system with $2 \times 2$ unit cells and 12 sublattice sites, where $(J_1,J_2,J_{\chi},J_r)=(1,1.4,0.7,0.3), (1,0.9,1,0.5), (1,1.9,2.72,1.6), (1,1.1,2.4,1.2)$ for the NOQF, FM, NACSL and the Trimerized respectively.}
	\begin{ruledtabular}
	\begin{tabular}{ccccc}
	Phase  & NOQF  & FM  & NACSL  & Trimerized\\
	\hline
	$E_1$           & -0.6524     & 2      & -0.9619        & -0.8386\\
	$E_2$          & 4.6074       & 2      & 3.6572          & 3.2917\\
	$E_{\chi}$   & 0.0058       & 0      & 0.5836          & 0.7429\\
	$E_r$          & -0.0170       & 0      & -0.4818        & -0.4818\\
	$\mathrm{CTB}_{(\bigtriangleup,\bigtriangledown)}$        & -0.0005      & 0       & 0.4652         & 0.3344\\
	$\mathrm{CTB}_{(\bigtriangleup,\bigtriangleup)}$        & 0.0092        & 0       & 0.7656         & 1.1111\\
	\end{tabular}
\end{ruledtabular}
\end{table*}

The phase diagram in Fig.~\ref{fig2}(a) is qualitatively consistent with the results of a small-sized system with $2 \times 2$ unit cells obtained by exact diagonalization (ED). The nature of the ED ground state can be read out from $(E_1,E_2,E_{\chi},E_r)$, the average values of each term in the Hamiltonian Eq.~(\ref{eqn:spinmodel2}), namely
\begin{eqnarray}
E_1 &=&\sum_{\langle i,j\rangle}\langle \bm{\mathrm{S}}_i\cdot\bm{\mathrm{S}}_j \rangle/N,
\nonumber \\ 
E_2 &=& \sum_{\langle i,j\rangle}\langle (\bm{\mathrm{S}}_i\cdot\bm{\mathrm{S}}_j)^2 \rangle/N,
\nonumber \\ 
E_{\chi} &=& \sum_{\bigtriangleup,\bigtriangledown}\sin{\phi_{ijk}}\langle(\bm{\mathrm{S}}_i\times\bm{\mathrm{S}}_j)\cdot\bm{\mathrm{S}}_k\rangle/N,
\nonumber \\ 
E_r &=&\sum_{\bigtriangleup,\bigtriangledown}\langle R_{ijk} \rangle/N
\end{eqnarray}
with $N=12$. For instance, the pure nematic state and the FM state are characterized as $(E_1,E_2,E_{\chi},E_r)=(0,4,0,0)$ and $(2,2,0,0)$, respectively. When the three-body interaction coefficients $J_r$ and $J_{\chi}$ are both small, the ED results are $(E_1,E_2,E_{\chi},E_r)=(-0.6524,4.6074,0.0058,-0.017)$ for $J_2=1.4$, and $(2,2,0,0)$ for $J_2 = 0.9$ (see Tab.~\ref{ED}), suggesting that the system is in the NOQF and FM phases, respectively. In both cases, the three-body interaction terms $E_r$ and $E_{\chi}$ are strongly suppressed as expected. 
 	
When $J_r$ and $J_{\chi}$ are large enough, $E_r$ and $E_{\chi}$ become significant. At this time, the three low-lying states all have $S_{\rm total} = 0$ and are nearly degenerate, with a relatively large gap away from the fourth level, as shown in Fig.~\ref{fig2}(b). These results indicate a NACSL phase. In addition, we find a trimerized phase which spontaneously breaks the inversion symmetry of the lattice. This signature can be captured by the correlation of three-body terms $\mathrm{CTB}_{(ijk,mnl)}\equiv \langle R_{ijk}  \cdot R_{mnl}\rangle$ between two adjacent triangles $(ijk)$ and $(mnl)$. As shown in Tab.~\ref{ED}, in the trimerized phase the correlation $\mathrm{CTB}_{(\bigtriangleup,\bigtriangledown)}$ between a triangle $\bigtriangleup$ and its neighboring inverse-triangles $\bigtriangledown$ is very different from $\mathrm{CTB}_{(\bigtriangleup,\bigtriangleup)}$. Since the trimerized ground state has $S_{\rm total}=0$, we identify it as the trimerized $Z_2$-SL in the phase diagram Fig.~\ref{fig2}(a).

\section{Abelian and non-Abelian QSLs in cold spinor Bose gas}
\label{sec:QSL}
While the phase diagram of the effective spin model Fig.~\ref{fig2}(a) exhibits the exciting possibility of a NACSL phase, we also need to consider whether NACSL can still exist under the constraints of the physical parameters, since the spin model is only an approximation of the original Hubbard model Eq.~(\ref{eqn:spinmodel}) in the limit of weak hopping and single particle occupation. Thanks to the remarkable controllability of ultracold atoms, one can change the hopping rate $t$ by the lattice potential, and tuning the interaction ratio $U_2/U_0$ by via optical Feshbach resonance~\cite{Chin2010} or microwave-induced Feshbach resonance~\cite{Papoular2010}) in a large parameter range.

\begin{figure}[tbp]
	\centering
	\includegraphics[width=0.9\linewidth]{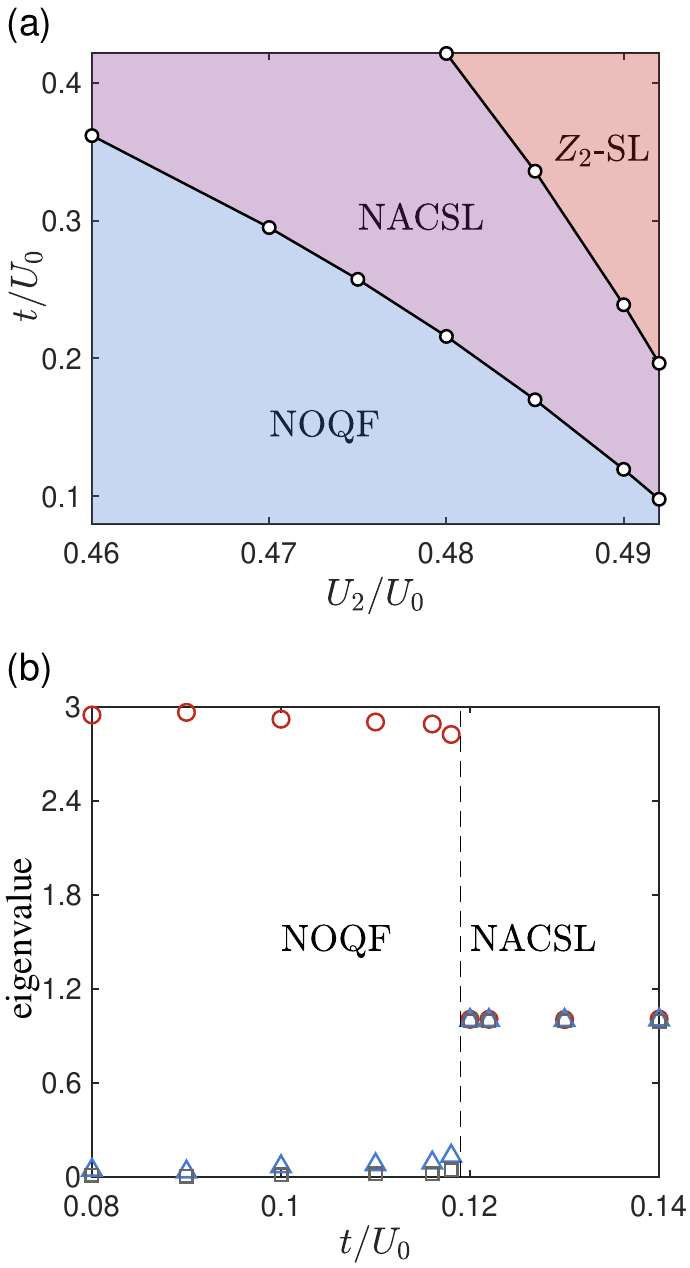}
	\caption{(a) The phase diagram of the Hubbard model Eq.~(\ref{eqn:spinmodel}) under the condition of weak hopping. (b) The  `density' weights of the three eigenstates are obtained by diagonalizing the fidelity matrix. The left region is a topological trivial NOQF with $D\approx2$, where the maximum eigenvalue is almost $3$. The right region is a topological NACSL, and three equally weighted eigenvalues $\sim1$ represent the triple degeneracy of the ground state. Here, $L_x=L_y=10$, $U_{2}/U_0=0.49$, $t^3/U^2_{0,2}<0.012$ with maximum hopping $t_{\rm max}/U_0=0.14$.}
	\label{fig3} 
\end{figure}

From the experimental point of view, it is more instructive to plot the phase diagram using the parameters of the Hubbard model Eq~(\ref{eqn:spinmodel}). To this end, we restrict in the small hopping regime and map out the phase diagram of the Hubbard model with the parameters $t/U_0$ and $U_2/U_0$. The result is shown in Fig.~\ref{fig3}(a), where the ratio $t/U_0$ is set to be less than 0.42, and the ratio of interaction $U_2/U_0$ is adopted with in the range from 0.46 to 0.49.


When the hopping $t$ is small, the system undergoes spontaneous symmetry breaking and the ground state exhibits NOQF. As $t$ increases, a NACSL phase shows up in which the three species of slave fermions $(c_1, c_{0}, c_{-1})$ each carries a Chern number $1$, leading to a total Chern number $\nu=3$.  A vortex excitation contains three Majorana zero modes (after projection the zero modes correspond to the Ising-type non-Abelian anyon) and hence obeys non-Abelian statistics. The  NACSL has a topological order described by an ${\rm SO}(3)_1$ Chern-Simons theory~\cite{Liu2018} and should have three degenerate ground states when defined on a torus. Thus, we insert $Z_2$ gauge fluxes by changing the boundary conditions from periodic to antiperiodic to obtain four different mean-field states: ($0,0$), ($0,\pi$), ($\pi,0$), and ($\pi,\pi$). Since ($0,0$) vanishes after Gutzwiller projection due to the incorrect fermi parity, we only consider the fidelity matrix $\rho$ for the Gutzwiller-projected states $\ket{G(0,\pi)}$, $\ket{G(\pi,0)}$, and $\ket{G(\pi,\pi)}$. The eigenvalues of $\rho$ stand for the `density' weights of the corresponding eigenstates and hence reflect the ground-state degeneracy. By diagonalizing $\rho$, we confirm the 3-fold ground-state degeneracy of the NACSL phase [Fig.~\ref{fig3}(b)]. 

Besides, a “trimerized” Abelian $Z_2$-SL with zero Chern number $\nu=0$ is found in the phase diagram Fig.~\ref{fig3}(a). The variational result of this phase features a large value of $T$, indicating strong `trimerization' and the breaking of inversion symmetry. Although strongly trimerized, this state is also topologically non-trivial because of a four-fold ground-state degeneracy on a torus. It seems to be a contradiction that the ground state is both symmetry breaking and topologically ordered. To understand this, we notice that the mean-field ground state is a BCS state where fermions form singlet pairs with total spin $S_{\rm total}=0$ (states with $S_{\rm total}\neq0$ are energetically unfavored). Although entanglement within the strong bond is stronger than the weak bond, only two of the three spins in each trimerized triangle can locally form a singlet-pair, and the third one must pair with another spin from other trimers. In other words, entanglement between different trimers is unavoidable given that $S_{\rm total}=0$. This can result in long-range entanglement and the topological degeneracy. Thus, the trimerized $Z_2$-SL is quite different from a valence-bond solid state where the three spins on a trimerized triangle locally form a three-body singlet~\cite{Liu2015,Cai2009}. 

\begin{figure}[tbp]
	\centering
	\includegraphics[width=1\linewidth]{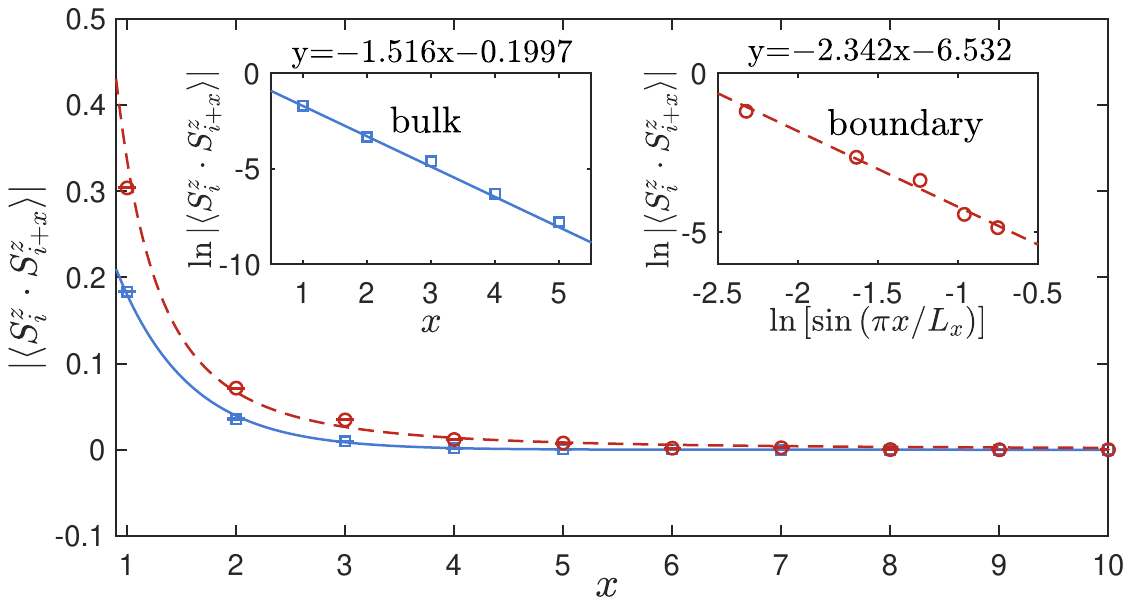}
	\caption{The $S_z$ component spin correlation of the NACSL phase on a cylinder with $16\times10$ unit cells, where the parameters are $(t,U_2)=(0.12,0.49)U_0$. The $x$-direction (16 unit cells) has a periodic boundary condition and the $y$-direction (10 unit cells) is open. The blue squares (red circles) represent the Monte Carlo data in the bulk (at the boundary), while the corresponding exponential (power law) fit are depicted by solid (dashed) lines. Inset: The left panel depicts a linear-log fit and the right panel shows a log-log fit.}
	\label{fig4} 
\end{figure}

Since the NACSL is of great importance, next we provide a method to detect it. Owing to the non-zero Chern number, the energy spectrum on the edge contains three branches of chiral Majorana excitations which remain gapless even after the Gutzwiller projection. The behavior of correlations of these gapless edge modes is hence different from that of the fully gapped bulk. In Fig.~\ref{fig4}, we demonstrate the spin-spin correlation for the $S_z$ component, which is exponentially decayed with distance in the bulk and power-law decayed at the boundary. This distinctive behavior of nontrivial edge spectrum may be detected experimentally in cold atomic gases by {\it in-situ} measurement of particle numbers in different spin states~\cite{Boll2016}.

Besides, one can also detect the topological orders in the QSL phases by extracting their nontrivial topological entanglement entropy~\cite{Kitaev2006L}, which can be obtained from the area law of second Renyi entanglement entropy~\cite{Zhang2011L}. While the measurement of entanglement entropy is elusive in solid state systems, in cold-atom platforms this can be done by interfering two identical copies and measuring by quantum gas microscopy with a single-site resolution~\cite{Islam2015}.

%


\section{Concluding remarks}
\label{sec:summary}

In summary, we propose to realize non-Abelian quantum spin liquid phases in a spinor BEC of ultracold atoms loaded on optical Kagome lattices. Using the technique of Raman coupling, a complex hopping is tuned to establish non-zero flux configuration in staggered lattice potential. When the hopping integral is weak, the system can be well approximated by an effective spin model of $S=1$, which supports a non-Abelian chiral spin liquid phase, along with an Abelian quantum spin liquid phase with symmetry breaking order in experimentally feasible parameter regions. The non-Abelian phase can be detected from the characteristic gapless edge states, and has potential applications in topological quantum computation. In addition to the experimental scheme illustrated in Fig.~\ref{fig1}, our proposal can also be implemented using stroboscopic optical tweezer arrays~\cite{Yan2022}, which can construct systems of nearly arbitrary geometry with independent control of the depths and positions of all sites. The tweezer arrays also facilitate the preparation and detection of various quantum states~\cite{Kaufman2021}. 

\begin{figure}[tbp]
	\centering
	\includegraphics[width=1\linewidth]{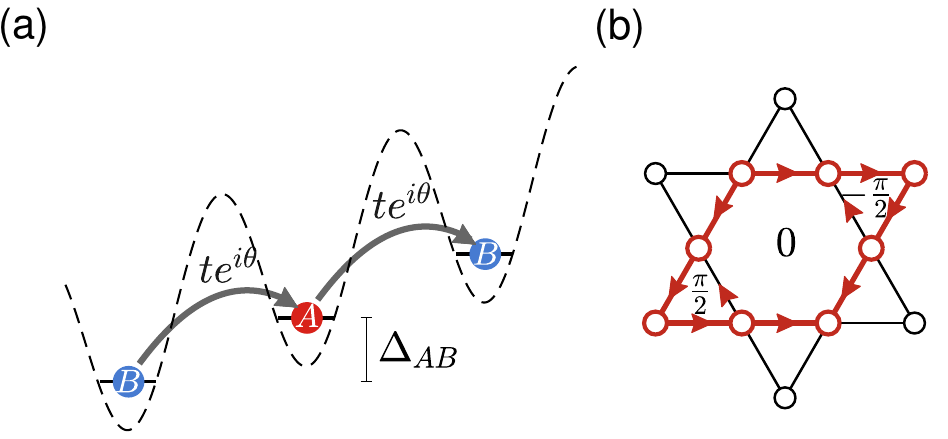}
	\caption{(a) A linearly tilted potential along the AB direction can generate a pattern with opposite fluxes in triangles and the inverted triangles shown in (b).}
	\label{fig5} 
\end{figure}

Finally, we briefly discuss another experimental setup for generating non-zero fluxes by changing the lattice potential. We replace the staggered LWW$_1$ and LWW$_2$ lattices by a linearly tiled potential, which can be achieved by accelerating the optical lattice~\cite{Jaksch2003}, and tuning the Raman tunnelings along the same direction to assume a same phase, as depicted in Fig.~\ref{fig5}(a). Different from the Kagome pattern [Fig.~\ref{fig1}(c)] with the same flux in the triangles and inverted triangles as discussed above, the new scheme results in a different pattern [Fig.~\ref{fig5}(b)] where the triangles and inverse triangles have opposite fluxes. This configuration can lead to more abundant Abelian QSLs, which break the SO(3) symmetry and have higher ground state degeneracy, as detailed in Appendix~\ref{app:SD}. However, the most interesting NACSL is not found in that case.

\acknowledgments
We thank Kuiyi Gao and Wei Zhu for helpful discussion, and financial support from the National Natural Science Foundation of China (Grant No.~12074428, No.~92265208, No.~11974421, and No.~12134020), and the National Key R\&D Program of China (Grant No.~2018YFA0306501 and No.~2022YFA1405301). Computational resources have been provided by the Physical Laboratory of High Performance Computing at Renmin University of China.

\appendix
\section{Raman coupling}
\label{app:Raman}
For cold atom systems, a complex tunneling can be induced by Raman coupling~\cite{Aidelsburger2011,Jaksch2003} or periodically modulation of lattice position~\cite{Jotzu2014}. Here, we focus on the Raman coupling technique and introduce a scheme to realize controllable complex tunneling coefficients.

First, we apply another two triangle lattices with long wavelength (LWW$_1$ and LWW$_2$) to lift the degeneracy of the three sublattices (A, B, and C) of the Kagome lattice and create an additional site-dependent potential $\mathcal{H}_{\rm add}=\sum_i \delta\epsilon_{i}\hat{n}_i$, as depicted in Fig.~\ref{fig1}(c) of the main text. There corresponding potential minima are located at sites A and B, respectively, so that the two sites can have an energy shift $\Delta_{AB}=|\delta\epsilon_A-\delta\epsilon_B|$. The energy offsets $\Delta_{AB}$ and $\Delta_{BC}$ can be freely adjusted by changing the light intensity of the LWW$_1$ and LWW$_2$ lasers. As a result, an additional site-dependent  potential is induced.

Then, we drive Raman coupling between nearest neighbors to realize a hopping with phase. Owing to the energy offsets, the Raman processes between different sublattice sites can be tuned independently. For instance, by applying a pair of Raman lasers along the AB direction with frequencies $\omega_{AB}$ and $\omega'_{AB}$, we can respectively drive a transition for atoms on the $\ket{A}$ and $\ket{B}$ sites to an excited state $\ket{e}$ with Rabi frequencies $\Omega_{Ae}$ and $\Omega'_{Be}$. The two running waves are linearly polarized with the same polarization direction and do not cause any transfer of angular momentum. In the presence of a large one-photon detuning $|\Delta^{\rm Raman}_{AB}| \gg \{ \Omega_{Ae}, \Omega'_{Be} \}$ and a small two-photon detuning $\delta_{AB}$, the excited state $\ket{e}$ can be adiabatically eliminated, and the effective Rabi frequency for the Raman transition between $\ket{A}$ and $\ket{B}$ reads
\begin{eqnarray}
	\label{Rb}
	\Omega_{AB}&=&\frac{\Omega_{Ae}\Omega'_{Be}}{2\Delta^{\rm Raman}_{AB}}
	\exp\left\{i \left[ \left(\mathbf{k}_{AB}-\mathbf{k}'_{AB}\right) \mathbf{r}+\left(\mathrm{\phi}_{AB}-\mathrm{\phi}'_{AB}\right) \right] \right\}\nonumber \\
	&& =\Omega^*_{BA   },
\end{eqnarray}
where $\mathrm{\phi_{AB}}$ and $\mathrm{\phi'_{AB}}$ are the initial phases of the two lasers, and $\mathbf{k}_{AB}$ and $\mathbf{k}'_{AB}$ are the corresponding wavevectors. The Raman processes between AC and BC sites can be implemented by another two pairs of Raman lasers analogously. The Hamiltonian describing Raman tunneling takes the form
\begin{eqnarray}
	\label{RM}
	\mathcal{H}_{\rm Raman}&=&-\sum_{\langle i,j\rangle,\sigma} \left(t_{ij}a_{\sigma j}^{\dagger}a_{\sigma i}+t_{ji}a^{\dagger}_{\sigma i}a_{\sigma j}\right) \nonumber \\
	&&+ \sum_i\delta_ia_{\sigma i}^{\dagger}a_{\sigma i},
\end{eqnarray}
where $\delta_i=\frac{\delta_{CA}-\delta_{AB}}{2}$ for $i\in$ $A$, $\frac{\delta_{AB}-\delta_{BC}}{2}$ for $i\in$ $B$, and $\frac{\delta_{BC}-\delta_{CA}}{2}$ for $i\in$ $C$. The Raman tunneling coefficient is 
\begin{eqnarray}
	\label{tso}
	t_{ij}=-\frac{1}{2}\int \mathrm{d}^2rw^*(\mathbf{r}-\mathbf{r}_i)w(\mathbf{r}-\mathbf{r}_j)\Omega_{ij}
\end{eqnarray}
with $w(\mathbf{r})$ the Wannier function. Owing to the explicit dependence on the effective Rabi frequency $\Omega_{ij}$, the modulus and phase of $t_{ij}$ can be tuned by properly choosing the intensity, wavelength, direction and initial phase of the Raman lasers. 

The total Hamiltonian becomes $\mathcal{H}=\mathcal{H}_{\rm latt}+\mathcal{H}_{\rm add}+\mathcal{H}_{\rm Raman}$. By adjusting the two-photon detuning to balance the offset of $\mathcal{H}_{\rm add}$, we get 
\begin{eqnarray}
	\label{tot}
	\mathcal{H}&=&\frac{U_0}{2}\sum_i\hat{n}_i(\hat{n}_i-1)-\sum_i\mu\hat{n}_i+ \frac{U_2}{2}\sum_i \left(\mathbf{S}^2_i-2\hat{n}_i \right) \nonumber \\
	&&-\sum_{\langle i,j\rangle,\sigma} \left(t_{ji}a_{\sigma i}^{\dagger}a_{\sigma j}+t_{ij}a^{\dagger}_{\sigma j}a_{\sigma i}\right).
\end{eqnarray}
For simplicity, we consider the case with $\Delta_{AB,BC,AC}\ll U_0$, and assume the same intensity and two-photon detuning of all Raman lasers. The initial phase difference of each pair of Raman beams is set as ${\pi}/{6}$, and the wave vectors satisfy $(\mathbf{k}-\mathbf{k}')\cdot \mathbf{r}_i=2\pi n$ with integer $n$. With that, we reach the desired condition of $|t_{ij}|=t$ and $|\theta_{ij}|= {\pi}/{6}$.

\section{Variational Monte Carlo approach}
\label{app:VMC}
\subsection{Aligned fluxes in triangles and inverse-triangles}

We first consider the configuration where the fluxes in triangles and inverse-triangles of the Kagome lattice are aligned, as shown in Fig.~\ref{fig1}(c). In this case, the natural choices~\cite{Liu2018} of variational parameters are $\chi_{ij}=\chi^{*}_{ji}$, $\Delta_{ij}=-\Delta_{ji}=\Delta e^{i\varphi_{ij}}$ (where $\varphi_{ij}$ is the angle between the bond and $x$-axis, depending on the bond and pairing symmetry), and $\lambda_i=\lambda$ playing the role of ``chemical potential". The mean-field Hamiltonian becomes
\begin{eqnarray}
	\label{mfhs}
	\mathcal{H}^{\mathrm{qsl}}_{\mathrm{MF}} &=& \sum_{\langle i,j \rangle} \left[\left(\chi C^{\dagger}_{ i}C_{ j}+\Delta e^{i\varphi_{ij}}C^{\dagger}_{ i}\bar{C}_{ j}\right)+\mathrm{H.c.}\right]\nonumber \\
	&& +\sum_i \lambda C^{\dagger}_{\alpha i}C_{\alpha i}.
\end{eqnarray}
It is useful to define a set of creation operators in Cartesian basis as linearly combinations of spinon operators
\begin{equation}
	\label{eq3}
	c^{\dagger}_x=\frac{1}{\sqrt{2}}(c^{\dagger}_{-1}-c^{\dagger}_1), \quad
	c^{\dagger}_y=\frac{i}{\sqrt{2}}(c^{\dagger}_{-1}+c^{\dagger}_1), \quad 
	c^{\dagger}_z=c^{\dagger}_0.
\end{equation}
With that, the hopping operator and singlet pairing operator can be written as 
\begin{eqnarray}
	\label{eq4}
	&&\hat{\chi}_{ij}=c^{\dagger}_{xi}c_{xj}+c^{\dagger}_{yi}c_{yj}+c^{\dagger}_{zi}c_{zj},\nonumber \\
	&&\hat{\Delta}_{ij}=-\left(c_{xi}c_{xj}+c_{yi}c_{yj}+c_{zi}c_{zj}\right).
\end{eqnarray}
And the mean-field Hamiltonian can be decomposed as $\mathcal{H}^{\mathrm{qsl}}_{\mathrm{MF}}=\sum_{s}\mathcal{H}^{s}_{\mathrm{MF}}$ showing an explicit SO(3) symmetry, where $s=x,y,z$, and 
\begin{eqnarray}
	\label{mf}
	{\cal H}^{s}_{\mathrm{MF}}&=&\sum_{\langle i,j \rangle} \left[\left(\chi c^{\dagger}_{s i}c_{s j}- \Delta e^{i\varphi_{ij}}c^{\dagger}_{s i}c^{\dagger}_{s j}\right)+\mathrm{H.c.}\right] \nonumber \\
	&&+  \sum_i \lambda c^{\dagger}_{s i}c_{s i}.
\end{eqnarray}

The ground state of the above Hamiltonian is a superconductor with ($p_x + ip_y$) symmetry, which contains two phases, i.e., a topological weak pairing phase and a trivial strong pairing phase~\cite{Read2000}. In order to ensure that each site is singly occupied, we need to perform the Gutzwiller projection on the mean-field wave function. Interestingly, the superconducting state before projection does not carry any intrinsic topological order, but after projection, the ground state wave function will carry an Abelian or non-Abelian topological order, depending on the topology of the mean-field ground state~\cite{Liu2018}. Specifically, when $-2|\chi|<\lambda<4|\chi|$, the superconductor is a weak pairing phase with non-trivial topology, which becomes a spin liquid phase with non-Abelian topological order after Gutzwiller projection. Otherwise, the superconductor is a topologically trivial strong pairing phase, and after projection becomes a spin liquid phase with $Z_2$ topological order.

We also consider various magnetically ordered phases as competing candidates, including lattice symmetry breaking orders where the triangle and inverse-triangle have different bond strengths, and nematic and ferromagnetic orders~\cite{Imambekov2003}. To this aim, we express the mean-field Hamiltonian as
\begin{eqnarray}
	\label{mean-field}
	\mathcal{H}_{\mathrm{MF}}&=&\sum_{\langle i,j \rangle} \left[ \left(\chi_{ij}C^{\dagger}_{ i}C_{ j}+\Delta_{ij} e^{i\varphi_{ij}}C^{\dagger}_{ i}\bar{C}_{j}\right)+\mathrm{H.c.}\right] +\nonumber \\
	&& \sum_{i} \lambda C^{\dagger}_{i}C_{ i} + D\sum_i (S^z_i)^2 + M \sum_i S^z_i,
\end{eqnarray}
where $T=\chi_{ij}/\chi_{kl}=\Delta_{ij}/\Delta_{kl}$ represents the ratio of the bond strengths between the triangle and inverse-triangle with $(i,j)$ denoteing sites belong to triangles and $(k,l)$ to inverse-triangles, and $M$ and $D$ are the background fields inducing the ferromagnetic and nematic orders, respectively. The other spin liquid terms, i.e. $\lambda,\Delta$ and $\chi$, govern the quantum fluctuations. To simplify notation, we denote the ground state of the above mean-field Hamiltonian Eq.~(\ref{mean-field}) as $\ket{\mathrm{MF}}$. With such notation, the trial wave function after Guzwiller projection $P_G$ reads $\ket{\Phi(\alpha)}=P_G\ket{\rm MF}$, where $\alpha=(\chi,\Delta,\lambda,D,M,T)$ are the variational parameters about to be determined by minimizing the energy functional $E(\alpha)=\bra{\Phi(\alpha)}{{\cal H}_{\mathrm{eff}}}\ket{\Phi(\alpha)}/\langle \Phi(\alpha)|  \Phi(\alpha)\rangle$. It is worth noting that if the ($p_x+ip_y$)-pairing is changed to ($p_x-ip_y$)-pairing, the energies associated with the three-body interaction will reverse sign to give $E_{\chi}\rightarrow -E_{\chi}$ and $E_r\rightarrow -E_r$. Therefore, we need to reasonably choose the pairing form, so that the three-body interaction can reduce the ground-state energy. For the case of $J_{\chi}>0$ and $|J_{\chi}|>|J_r|$ considered here, the ($p_x-ip_y$) pairing state is the one that fill the purpose.

\subsection{Opposite fluxes in triangles and inverse-triangles}

To study the configuration where the triangles and inverse-triangles have opposite fluxes as shown in Fig.~\ref{fig5}(b), we notice from Eq.~(\ref{eqn:spinmodel2}) that the value of flux $\phi_{ijk}$ affects the sign of the coefficients of $(\mathbf{S}_i\times \mathbf{S}_j)\cdot \mathbf{S}_k$ and $R_{ijk}$. For the case with aligned fluxes, the coefficients of the three-body interaction terms [$(\mathbf{S}_i\times \mathbf{S}_j)\cdot \mathbf{S}_k$ and $R_{ijk}$] are the same. For the case with opposite fluxes, the coefficients of three-body interaction terms of triangles and inverse-triangles are opposite. In addition, one can also conclude from Eqs.~(\ref{M:mean}) and (\ref{hr}) that if the hopping and pairing terms of a bond in a triangle (or an inverse-triangle) change sign, the flux therein will flip. Thus, we treat this configuration by imposing an additional minus sign onto the hopping and pairing terms of a bond in triangles (or inverse-triangles) based on the mean-field Hamiltonian Eq.~(\ref{mean-field}) for the case with aligned fluxes. The following analysis is then carried analogously.

\begin{table*}
	\caption{\label{SMK}Variational Monte Carlo results for the case of aligned fluxes. The parameters $(J_2,J_{\chi})$ are $(1.3,2)$ for NOQF, $(0.9,2)$ for FM, $(1.5,3.28)$ for NACSL, and $(1,2.6)$ for $Z_2$-SL phases. Other parameters are $J_1 = 1$, $J_r=0.5$, $\chi = -1$, and $L_x=L_y=6$.} 
	\begin{ruledtabular}
		\begin{tabular}{cccccccccccc}
			&$\Delta$ &$\lambda$ &$D$ &$T$ & $M$ & $E_1$  & $E_2$  & $E_{\chi}$  & $E_r$   & $E$ &($\nu_0,\nu_1,\nu_{-1}$)\\
			\hline
			NOQF     &5.065 &-8.821 &14.544 &1.016 &0.315  & -0.422     & 4.374      & $< 10^{-3}$     &0.001     & -5.265  &/ \\
			FM        &1.2 &0.8 &0.01 &1.0 &15 & 2       & 2      & 0      & 0      & 3.8  &/\\
			NACSL    &1.858 &0.429 &0.170 &1.056 &0.549  & -1.133     & 3.692      & 0.477        &-0.369    & -6.155    & (-1,-1,-1)\\
			$Z_2$-SL   &0.845 &8.979 &2.441 &258.253 &-0.177  & -0.731      & 3.033     & 0.580    & -0.574      & -4.097   &(0,0,0)\\
		\end{tabular}
	\end{ruledtabular}
\end{table*}
%

\section{Supplemental data for the effective spin-model}
\label{app:SD}
\subsection{VMC results for the case with aligned fluxes}

In Table.~\ref{SMK}, we show the results of the variational Monte Carlo approach for the above four phases with typical choices of parameters. For the FM phase, the parameter $M$ is dominant over the other parameters, representing a complete polarization of all spins. For the NOQF phase, the nematic order parameter $D$ is much larger than all other variational coefficients, with a ground-state energy less than $-4J_2$. This result indicates that the phase is not a pure nematic phase (which has $E_1=0$, $E_2=4J_2$, and $E=-4J_2$), but a nematic phase with quantum fluctuations. For the NACSL phase, we get $D\approx0$, $M\approx0$, $T\approx1$, and $\lambda<4$, suggesting a weak pairing phase as the ground state, which will carry a non-Abelian topological order after projection. For the $Z_2$-SL phase, we obtain $D\approx0$, $M\approx0$, $\lambda>>4$, and a very large value of $T$. This result indicates that the mean-field ground state is a strong pairing state, which carries a $Z_2$ topological order after projection. Nevertheless, we find that the huge $T$ does not make the ground state a trivial trimer (TT) where each triangle is independent, but a trimerized $Z_2$-SL which is still topologically degenerate after projection.

In order to further characterize the trimerized $Z_2$-SL, we calculate the ground state of the TT through exact diagonalization. Owing to the lack of long-range resonances, the ground state energy of the TT can be simply obtained by diagonalizing a three-particle Hamiltonian. We choose the parameters of the spin model as $(J_1,J_2,J_{\chi},J_r)=(1,1.5,3,2.5)$. The result of diagonalization shows that the ground state of a triangle acquires triple degeneracy with trimerized total spin $S_t=1$, which is different from the single state trimer with $S_t=0$~\cite{Liu2015,Cai2009}. This spin quantum number of  TT is obviously inconsistent with our variational result giving total spin $0$. For this reason, we calculate the average expectation of energies $(E_1,E_2,E_{\chi},E_r)$ of the two ground states [see the main text for the definitions of $(E_1,E_2,E_{\chi},E_r)$]. For the trivial TT, we have $(E_1,E_2,E_{\chi},E_r)=(-0.6667,1.6667,0.5774,-0.5774)$; for the variational ground state, we have $(E_1,E_2,E_{\chi},E_r)=(-0.7387,3.0393,0.5817,-0.5734)$. Notice that the three-body interaction energies ($E_{\chi}$ and $E_r$) of TT are similar to the variational ground state, but there is a considerable difference in the two-body interaction energies ($E_1$ and $E_2$). This observation suggests that in both possibilities, each triangle is a trimer state with $S_t=1$ and gives the same three-body interaction energies. However, the triangles are independent for the TT, while the triangles are resonant in the variational ground state since it is constrained by a total spin $0$. Thus, the two-body interaction energies in the latter case are greater than the TT.
\begin{table*}
	\caption{\label{CMK}Variational Monte Carlo results for the case of opposite fluxes. The parameters are $J_1=1$, $J_2=1.5$, $J_{\chi}=12$, $\chi = -1$, and $L_x=L_y=6$.}
	\begin{ruledtabular}
		\begin{tabular}{cccccccccccc}
			$J_r$	&$\Delta$ &$\lambda$ &$D$ &$T$ & $M$ & $E_1$  & $E_2$  & $E_{\chi}$  & $E_r$   & $E$ &($\nu_0,\nu_1,\nu_{-1}$)\\
			\hline
			-4     &0.933 &12.794 &-0.037 &17.910 &0.325  & -0.784     & 3.073     & 0.572     & -0.558     & -8.458  &(0,0,0) \\
			-5.2        &1.000 &3.009 &0.077 &2.005 &0.003 & -1.004       & 3.307     & 0.510      & -0.437   & -7.804  &(-6,0,0)\\
			-6    &0.083 &1.507 &0.011 &0.255 &0.004  & -1.095       & 3.477      & 0.405       & -0.236   & -7.565   & (0,0,0)\\
			-9.3   &0.016 &1.131 &0.032 &0.081 &0.001 & -1.064      & 3.405      & 0.365   &-0.159    & -6.945  &(-2,0,0)\\
			-9.8   &0.006 &1.020 &$<10^{-3}$ &0.012 &0.002 & -1.024      &3.283    & 0.316   & -0.086     & -6.875   &(-2,-2,-2)\\
			
		\end{tabular}
	\end{ruledtabular}
\end{table*}
%

\subsection{VMC results for the case with opposite fluxes}

\begin{figure}[tbp]
	\centering
	\includegraphics[width=1\linewidth]{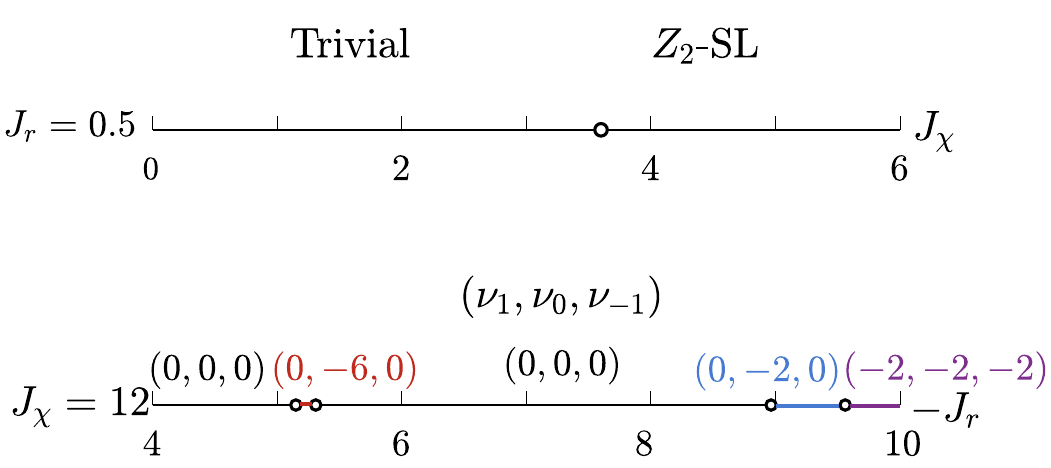}
	\caption{ The phase diagram with parameters $(J_1,J_2) = (1,1.5)$.
		For $J_r = 0.5$ (top), we observe the topological transition from the trivial phase to the $Z_2$-SL. For $J_{\chi} = 12$ (bottom), $J_r$ and $J_{\chi}$ compete with each other and various Abelian QSLs with Chern numbers $(\nu_1, \nu_0, \nu_{-1})$ of different spin components are observed.}
	\label{figs3} 
\end{figure}

For the configuration with opposite fluxes as shown in Fig.~\ref{fig5}(b), the effective spin-model preserves SO(3) spin rotation symmetry but breaks inversion symmetry of the lattice. The breaking of lattice inversion symmetry suggests that there may be more possible QSL phases. Indeed, we find various Abelian QSLs, as shown in Fig.~\ref{figs3}, with detailed numerical results listed in Table.~\ref{CMK}. These QSLs are distinguished by the Chern numbers ($\nu_1,\nu_0,\nu_{-1}$) of the three spin components. Interestingly, we find Abelian QSLs that spontaneously break the SO(3) symmetry, such as ($\nu_1,\nu_0,\nu_{-1}$)=($0,-6,0$) and ($0,-2,0$), which are not found in the aligned fluxes setup [Fig.~\ref{fig1}(c)].

It is worth noting that in this pattern, the effective Hamiltonian Eq.~(\ref{eqn:spinmodel2}) is invariant under the joint operation of time inversion and lattice inversion. When we select the mean-field ground state, an extra minus sign is introduced to the hopping coefficients on the bonds of the inverse-triangle, which explicitly breaks the symmetry and reduces the dimensionality of the ground state subspace. In the main text, we choose the $(p_x-ip_y)$ pairing state as the mean field and assign opposite fluxes in the triangles, leading to a mean field state denoted as $\ket{p_x-ip_y,d}$. Alternatively, we can also change the pairing symmetry to $(p_x+ip_y)$ and assign opposite fluxes in the inverse-triangles, and obtain a new mean field state $\ket{p_x+ip_y,u}$. Both choices will give an exactly same ground state energy. Taking a torus systems of size $6 \times 6$ as an example, we calculate the expectation values of different terms for a typical choice of paramters  $(\chi,\Delta,\lambda,D,T,M)=(-1,1,3.009,0.077,2.005,0.001)$. For the $P_G\ket{p_x-ip_y,d}$ state, we obtain $(E_1,E_2,E_{\chi},E_r) = (-1.003,3.306,0.509,-0.437)$, while the same results for the $P_G\ket{p_x+ip_y,u}$ state are  $(-1.003,3.305,0.509,-0.437)$. This finding means that the ground state degeneracy should be doubled by considering the symmtry.


\begin{thebibliography}{60}	

\bibitem{Thouless1982}
\bibinfo{author}{D.~J. Thouless}, \bibinfo{author}{M. Kohmoto}, \bibinfo{author}{M.~P. Nightingale} and \bibinfo{author}{M. den Nijs},
\newblock \bibinfo{title}{Quantized Hall Conductance in a Two-Dimensional Periodic Potential},
\href{https://journals.aps.org/prl/abstract/10.1103/PhysRevLett.49.405}{
\newblock \bibinfo{journal}{Phys. Rev. Lett.}
\textbf{\bibinfo{volume}{49}}, \bibinfo{pages}{405}
(\bibinfo{year}{1982})}.


\bibitem{Wen1990T}
\bibinfo{author}{X.-G. Wen},
\newblock \bibinfo{title}{Topological orders in rigid states},
\href{https://www.worldscientific.com/doi/abs/10.1142/S0217979290000139}{
\newblock \bibinfo{journal}{Int. J. Mod. Phys. B}
\textbf{\bibinfo{volume}{04}}, \bibinfo{pages}{239}
(\bibinfo{year}{1990})}.


\bibitem{Zhou2017}
\bibinfo{author}{Y. Zhou}, \bibinfo{author}{K. Kanoda}, and \bibinfo{author}{T.-K. Ng},
\newblock \bibinfo{title}{Quantum spin liquid states},
\href{https://journals.aps.org/rmp/abstract/10.1103/RevModPhys.89.025003}{
\newblock \bibinfo{journal}{Rev. Mod. Phys.}
\textbf{\bibinfo{volume}{89}}, \bibinfo{pages}{025003}
(\bibinfo{year}{2017})}.


\bibitem{Anderson1987}
\bibinfo{author}{P.~W. Anderson}, 
\newblock \bibinfo{title}{The Resonating Valence Bond State in La$_2$CuO$_4$ and Superconductivity},
\href{https://www.science.org/doi/10.1126/science.235.4793.1196}{\newblock \bibinfo{journal}{Science}
\textbf{\bibinfo{volume}{235}}, \bibinfo{pages}{1196}
(\bibinfo{year}{1987})}.


\bibitem{Wen2017}
\bibinfo{author}{X.~G. Wen},
\newblock \bibinfo{title}{Colloquium: Zoo of quantum-topological phases of matter},
\href{https://journals.aps.org/rmp/abstract/10.1103/RevModPhys.89.041004}{
\newblock \bibinfo{journal}{Rev. Mod. Phys.}
\textbf{\bibinfo{volume}{89}}, \bibinfo{pages}{041004}
(\bibinfo{year}{2017})}.


\bibitem{Wen1989}
\bibinfo{author}{X.~G. Wen},
\newblock \bibinfo{title}{Vacuum degeneracy of chiral spin states in compactified spaces},
\href{https://journals.aps.org/prb/abstract/10.1103/PhysRevB.40.7387}{
\newblock \bibinfo{journal}{Phys. Rev. B}
\textbf{\bibinfo{volume}{40}}, \bibinfo{pages}{7387}
(\bibinfo{year}{1989})}.	


\bibitem{Wen1990}
\bibinfo{author}{X.~G. Wen} and \bibinfo{author}{Q. Niu},
\newblock \bibinfo{title}{Ground-state degeneracy of the fractional quantum Hall states in the presence of a random potential and on high-genus Riemann surfaces},
\href{https://journals.aps.org/prb/abstract/10.1103/PhysRevB.41.9377}{
\newblock \bibinfo{journal}{Phys. Rev. B}
\textbf{\bibinfo{volume}{41}}, \bibinfo{pages}{9377}
(\bibinfo{year}{1990})}.	


\bibitem{Chen2010}
\bibinfo{author}{X. Chen}, \bibinfo{author}{Z.-C. Gu} and \bibinfo{author}{X.-G. Wen},
\newblock \bibinfo{title}{Local unitary transformation, long-range quantum entanglement, wave function renormalization, and topological order},
\href{https://journals.aps.org/prb/abstract/10.1103/PhysRevB.82.155138}{
\newblock \bibinfo{journal}{Phys. Rev. B}
\textbf{\bibinfo{volume}{82}}, \bibinfo{pages}{155138}
(\bibinfo{year}{2010})}.


\bibitem{Kitaev2006}
\bibinfo{author}{A. Kitaev}, 
\newblock \bibinfo{title}{Anyons in an exactly solved model and beyond},
\href{https://www.sciencedirect.com/science/article/pii/S0003491605002381?via\%3Dihub}{
\newblock \bibinfo{journal}{Ann. Phys. (Amsterdam)}
\textbf{\bibinfo{volume}{321}}, \bibinfo{pages}{2}
(\bibinfo{year}{2006})}.


\bibitem{Greiter2009}
\bibinfo{author}{M. Greiter} and \bibinfo{author}{R. Thomale},
\newblock \bibinfo{title}{Non-Abelian Statistics in a Quantum Antiferromagnet},
\href{https://journals.aps.org/prl/abstract/10.1103/PhysRevLett.102.207203}{
\newblock \bibinfo{journal}{Phys. Rev. Lett.}
\textbf{\bibinfo{volume}{102}}, \bibinfo{pages}{207203}
(\bibinfo{year}{2009})}.


\bibitem{Nayak2008}
\bibinfo{author}{C. Nayak}, \bibinfo{author}{S.~H. Simon}, \bibinfo{author}{A. Stern}, \bibinfo{author}{M. Freedman}, and \bibinfo{author}{S.~D. Sarma},
\newblock \bibinfo{title}{Non-Abelian Anyons and Topological Quantum Computation},
\href{https://journals.aps.org/rmp/abstract/10.1103/RevModPhys.80.1083}{
\newblock \bibinfo{journal}{Rev. Mod. Phys.}
\textbf{\bibinfo{volume}{80}}, \bibinfo{pages}{1083}
(\bibinfo{year}{2008})}.


\bibitem{Savary2017}
\bibinfo{author}{L. Savary} and \bibinfo{author}{L. Balents},
\newblock \bibinfo{title}{Quantum spin liquids: a review},
\href{https://iopscience.iop.org/article/10.1088/0034-4885/80/1/016502}{
\newblock \bibinfo{journal}{Rep. Prog. Phys.}
\textbf{\bibinfo{volume}{80}}, \bibinfo{pages}{016502}
(\bibinfo{year}{2017})}.


\bibitem{Ran2007}
\bibinfo{author}{Y. Ran}, \bibinfo{author}{M. Hermele}, \bibinfo{author}{P. A. Lee}, and \bibinfo{author}{X.-G. Wen},
\newblock \bibinfo{title}{Projected-Wave-Function Study of the Spin-1/2 Heisenberg Model on the Kagom\'{e} Lattice},
\href{https://journals.aps.org/prl/abstract/10.1103/PhysRevLett.98.117205}{
\newblock \bibinfo{journal}{Phys. Rev. Lett.}
\textbf{\bibinfo{volume}{98}}, \bibinfo{pages}{117205}
(\bibinfo{year}{2007})}.


\bibitem{Han2012}
\bibinfo{author}{T.-H. Han}, \bibinfo{author}{J.~S. Helton}, \bibinfo{author}{S. Chu}, \bibinfo{author}{D.~G. Nocera}, \bibinfo{author}{J.~A. Rodriguez-Rivera}, \bibinfo{author}{C. Broholm}, and \bibinfo{author}{Y.~S. Lee},
\newblock \bibinfo{title}{Fractionalized excitations in the spin-liquid state of a kagome-lattice antiferromagnet}, \href{https://www.nature.com/articles/nature11659}{\newblock \bibinfo{journal}{Nature (London)}
\textbf{\bibinfo{volume}{492}}, \bibinfo{pages}{406}
(\bibinfo{year}{2012})}.


\bibitem{Cheng2011}
\bibinfo{author}{J.~G. Cheng}, \bibinfo{author}{G. Li}, \bibinfo{author}{L. Balicas}, \bibinfo{author}{J.~S. Zhou}, \bibinfo{author}{J.~B. Goodenough}, \bibinfo{author}{C. Xu}, and \bibinfo{author}{H.~D. Zhou},
\newblock \bibinfo{title}{High-Pressure Sequence Ba$_3$NiSb$_2$O$_9$ of Structural Phases: New $S=1$ Quantum Spin Liquids Based on Ni$^{2+}$},
\href{https://journals.aps.org/prl/abstract/10.1103/PhysRevLett.107.197204}{\newblock \bibinfo{journal}{Phys. Rev. Lett.}
\textbf{\bibinfo{volume}{107}}, \bibinfo{pages}{197204}
(\bibinfo{year}{2011})}.


\bibitem{Liu2018}
\bibinfo{author}{Z.-X. Liu}, \bibinfo{author}{H.-H. Tu}, \bibinfo{author}{Y.-H. Wu}, \bibinfo{author}{R.-Q. He}, \bibinfo{author}{X.-J. Liu}, \bibinfo{author}{Y. Zhou}, and \bibinfo{author}{T.-K. Ng},
\newblock \bibinfo{title}{Non-Abelian $S=1$ chiral spin liquid on the kagome lattice},
\href{https://journals.aps.org/prb/abstract/10.1103/PhysRevB.97.195158}{\newblock \bibinfo{journal}{Phys. Rev. B}
\textbf{\bibinfo{volume}{97}}, \bibinfo{pages}{195158}
(\bibinfo{year}{2018})}.


\bibitem{Fu2015}
\bibinfo{author}{M. Fu}, \bibinfo{author}{T. Imai}, \bibinfo{author}{T.-H. Han}, and \bibinfo{author}{Y.~S. Lee},
\newblock \bibinfo{title}{Evidence for a gapped spin-liquid ground state in a kagome Heisenberg antiferromagnet},
\href{https://www.science.org/doi/10.1126/science.aab2120}{\newblock \bibinfo{journal}{Science}
\textbf{\bibinfo{volume}{350}}, \bibinfo{pages}{655}
(\bibinfo{year}{2015})}.


\bibitem{Bloch2008}
\bibinfo{author}{I. Bloch}, \bibinfo{author}{J. Dalibard}, and \bibinfo{author}{W. Zwerger},
\newblock \bibinfo{title}{Many-body physics with ultracold gases},
\href{https://journals.aps.org/rmp/abstract/10.1103/RevModPhys.80.885}{
\newblock \bibinfo{journal}{Rev. Mod. Phys.}
\textbf{\bibinfo{volume}{80}}, \bibinfo{pages}{885}
(\bibinfo{year}{2008})}.


\bibitem{Eckardt2017}
\bibinfo{author}{A. Eckardt},
\newblock \bibinfo{title}{Colloquium: Atomic quantum gases in periodically driven optical lattices},
\href{https://journals.aps.org/rmp/abstract/10.1103/RevModPhys.89.011004}{
\newblock \bibinfo{journal}{Rev. Mod. Phys.}
\textbf{\bibinfo{volume}{89}}, \bibinfo{pages}{011004}
(\bibinfo{year}{2017})}.


\bibitem{Jotzu2014}
\bibinfo{author}{G. Jotzu}, \bibinfo{author}{M. Messer}, \bibinfo{author}{R. Desbuquois}, \bibinfo{author}{M. Lebrat}, \bibinfo{author}{T. Uehlinger}, \bibinfo{author}{D. Greif}, and \bibinfo{author}{T. Esslinger},
\newblock \bibinfo{title}{Experimental realization of the topological Haldane model with ultracold fermions}, \href{https://www.nature.com/articles/nature13915}{\newblock \bibinfo{journal}{Nature (London)}
\textbf{\bibinfo{volume}{515}}, \bibinfo{pages}{237}
(\bibinfo{year}{2014})}.


\bibitem{Song2018}
\bibinfo{author}{B. Song}, \bibinfo{author}{L. Zhang}, \bibinfo{author}{C. He}, \bibinfo{author}{T.~F.~J. Poon}, \bibinfo{author}{E. Hajiyev}, \bibinfo{author}{S. Zhang}, \bibinfo{author}{X.-J. Liu}, and \bibinfo{author}{G.-B. Jo},
\newblock \bibinfo{title}{Observation of symmetry-protected topological band with ultracold fermions},
\href{https://www.science.org/doi/10.1126/sciadv.aao4748}{
\newblock \bibinfo{journal}{Sci. Adv.}
\textbf{\bibinfo{volume}{4}}, \bibinfo{pages}{eaao4748}
(\bibinfo{year}{2018})}.	


\bibitem{Cooper2019}
\bibinfo{author}{N.~R. Cooper}, \bibinfo{author}{J. Dalibard}, and \bibinfo{author}{I.~B. Spielman},
\newblock \bibinfo{title}{Topological bands for ultracold atoms},
\href{https://journals.aps.org/rmp/abstract/10.1103/RevModPhys.91.015005}{
\newblock \bibinfo{journal}{Rev. Mod. Phys.}
\textbf{\bibinfo{volume}{91}}, \bibinfo{pages}{015005}
(\bibinfo{year}{2019})}.


\bibitem{Greiner-02}
\bibinfo{author}{M. Greiner}, \bibinfo{author}{O. Mandel}, \bibinfo{author}{T. Esslinger}, \bibinfo{author}{T.~W. H{\"a}nsch}, and \bibinfo{author}{I. Bloch},
\newblock \bibinfo{title}{Quantum phase transition from a superfluid to a Mott insulator in a gas of ultracold atoms},
\href{https://doi.org/10.1038/415039a}{
\newblock \bibinfo{journal}{Nature}
\textbf{\bibinfo{volume}{415}}, \bibinfo{pages}{39}
(\bibinfo{year}{2002})}.


\bibitem{Mazurenko2017}
\bibinfo{author}{A. Mazurenko}, \bibinfo{author}{C.~S. Chiu}, \bibinfo{author}{G. Ji}, \bibinfo{author}{M.~F. Parsons}, \bibinfo{author}{M. Kan\'{a}sz-Nagy}, \bibinfo{author}{R. Schmidt}, \bibinfo{author}{F. Grusdt}, \bibinfo{author}{E. Demler}, \bibinfo{author}{D. Greif}, and \bibinfo{author}{M. Greiner },
\newblock \bibinfo{title}{A cold-atom Fermi–Hubbard antiferromagnet}, \href{https://www.nature.com/articles/nature22362}{\newblock \bibinfo{journal}{Nature (London)}
\textbf{\bibinfo{volume}{545}}, \bibinfo{pages}{462}
(\bibinfo{year}{2017})}.


\bibitem{Goldman2016}
\bibinfo{author}{N. Goldman}, \bibinfo{author}{J.~C. Budich}, and \bibinfo{author}{P. Zoller},
\newblock \bibinfo{title}{Topological quantum matter with ultracold gases in optical lattices},
\href{https://www.nature.com/articles/nphys3803}{\newblock \bibinfo{journal}{Nat. Phys.}
\textbf{\bibinfo{volume}{12}}, \bibinfo{pages}{639}
(\bibinfo{year}{2016})}.


\bibitem{Semeghini2021}
\bibinfo{author}{G. Semeghini}, \bibinfo{author}{H. Levine}, \bibinfo{author}{A. Keesling}, \bibinfo{author}{S. Ebadi}, \bibinfo{author}{H. Levine}, \bibinfo{author}{T.~T. Wang}, \bibinfo{author}{D. Bluvstein}, \bibinfo{author}{R. Verresen}, \bibinfo{author}{H. Pichler}, \bibinfo{author}{M. Kalinowski}, \bibinfo{author}{R. Samajdar}, \bibinfo{author}{A. Omran}, \bibinfo{author}{S. Sachdev}, \bibinfo{author}{A. Vishwanath}, \bibinfo{author}{M. Greiner}, \bibinfo{author}{V. Vuleti\'{c}}, and \bibinfo{author}{M. D. Lukin},
\newblock \bibinfo{title}{Probing topological spin liquids on a programmable quantum simulator},
\href{https://www.science.org/doi/10.1126/science.abi8794}{\newblock \bibinfo{journal}{Science}
\textbf{\bibinfo{volume}{374}}, \bibinfo{pages}{1242}
(\bibinfo{year}{2021})}.


\bibitem{Giudici2022}
\bibinfo{author}{G. Giudici}, \bibinfo{author}{M.~D. Lukin}, and \bibinfo{author}{H. Pichler},
\newblock \bibinfo{title}{Dynamical Preparation of Quantum Spin Liquids in Rydberg Atom Arrays},
\href{https://journals.aps.org/prl/abstract/10.1103/PhysRevLett.129.090401}{
\newblock \bibinfo{journal}{Phys. Rev. Lett.}
\textbf{\bibinfo{volume}{129}}, \bibinfo{pages}{090401}
(\bibinfo{year}{2022})}.

\bibitem{Tarabunga2022}
\bibinfo{author}{P.~S. Tarabunga}, \bibinfo{author}{F.~M. Surace}, \bibinfo{author}{R. Andreoni}, \bibinfo{author}{A. Angelone}, and \bibinfo{author}{M. Dalmonte},
\newblock \bibinfo{title}{Gauge-Theoretic Origin of Rydberg Quantum Spin Liquids},
\href{https://journals.aps.org/prl/abstract/10.1103/PhysRevLett.129.195301}{
\newblock \bibinfo{journal}{Phys. Rev. Lett.}
\textbf{\bibinfo{volume}{129}}, \bibinfo{pages}{195301}
(\bibinfo{year}{2022})}.

\bibitem{Kalinowski2023}
\bibinfo{author}{M. Kalinowski}, \bibinfo{author}{N. Maskara}, and \bibinfo{author}{M.~D. Lukin},
\newblock \bibinfo{title}{Non-Abelian Floquet Spin Liquids in a Digital Rydberg Simulator},
\href{https://journals.aps.org/prx/abstract/10.1103/PhysRevX.13.031008}{
\newblock \bibinfo{journal}{Phys. Rev. X}
\textbf{\bibinfo{volume}{13}}, \bibinfo{pages}{031008}
(\bibinfo{year}{2023})}.


\bibitem{Sun2023}
\bibinfo{author}{B.-Y. Sun}, \bibinfo{author}{N. Goldman}, \bibinfo{author}{M. Aidelsburger}, and \bibinfo{author}{M. Bukov},
\newblock \bibinfo{title}{Engineering and Probing Non-Abelian Chiral Spin Liquids Using Periodically Driven Ultracold Atoms},
\href{https://journals.aps.org/prxquantum/abstract/10.1103/PRXQuantum.4.020329}{
\newblock \bibinfo{journal}{PRX Quantum}
\textbf{\bibinfo{volume}{4}}, \bibinfo{pages}{020329}
(\bibinfo{year}{2023})}.


\bibitem{Rudner2020}
\bibinfo{author}{M.S. Rudner} and \bibinfo{author}{N.H. Lindner},
\newblock \bibinfo{title}{Band structure engineering and non-equilibrium dynamics in Floquet topological insulators},
\href{https://www.nature.com/articles/s42254-020-0170-z}{
\newblock \bibinfo{journal}{Nat. Rev. Phys.}
\textbf{\bibinfo{volume}{2}}, \bibinfo{pages}{229}
(\bibinfo{year}{2020})}.


\bibitem{Zhao2015}
\bibinfo{author}{L. Zhao}, \bibinfo{author}{J. Jiang}, \bibinfo{author}{T. Tang}, \bibinfo{author}{M. Webb}, and \bibinfo{author}{Y. Liu},
\newblock \bibinfo{title}{Antiferromagnetic Spinor Condensates in a Two-Dimensional Optical Lattice},
\href{https://journals.aps.org/prl/abstract/10.1103/PhysRevLett.114.225302}{\newblock \bibinfo{journal}{Phys. Rev. Lett.}
\textbf{\bibinfo{volume}{114}}, \bibinfo{pages}{225302}
(\bibinfo{year}{2015})}.


\bibitem{Jo2012}
\bibinfo{author}{G.-B. Jo}, \bibinfo{author}{J. Guzman}, \bibinfo{author}{C. K. Thomas}, \bibinfo{author}{P. Hosur}, \bibinfo{author}{A. Vishwanath}, and \bibinfo{author}{D. M. Stamper-Kurn},
\newblock \bibinfo{title}{Ultracold Atoms in a Tunable Optical Kagome Lattice},
\href{https://journals.aps.org/prl/abstract/10.1103/PhysRevLett.108.045305}{
\newblock \bibinfo{journal}{Phys. Rev. Lett.}
\textbf{\bibinfo{volume}{108}}, \bibinfo{pages}{045305}
(\bibinfo{year}{2012})}.


\bibitem{Imambekov2003}
\bibinfo{author}{A. Imambekov}, \bibinfo{author}{M. Lukin}, and \bibinfo{author}{E. Demler},
\newblock \bibinfo{title}{Spin-exchange interactions of spin-one bosons in optical lattices: Singlet, nematic, and dimerized phases},
\href{https://journals.aps.org/pra/abstract/10.1103/PhysRevA.68.063602}{
\newblock \bibinfo{journal}{Phys. Rev. A}
\textbf{\bibinfo{volume}{68}}, \bibinfo{pages}{063602}
(\bibinfo{year}{2003})}.




\bibitem{Aidelsburger2011}
\bibinfo{author}{M. Aidelsburger}, \bibinfo{author}{M. Atala}, \bibinfo{author}{S. Nascimb\`{e}ne}, \bibinfo{author}{S. Trotzky}, \bibinfo{author}{Y.-A. Chen}, and \bibinfo{author}{I. Bloch},
\newblock \bibinfo{title}{Experimental Realization of Strong Effective Magnetic Fields in an Optical Lattice},
\href{https://journals.aps.org/prl/abstract/10.1103/PhysRevLett.107.255301}{
\newblock \bibinfo{journal}{Phys. Rev. Lett.}
\textbf{\bibinfo{volume}{107}}, \bibinfo{pages}{255301}
(\bibinfo{year}{2011})}.


\bibitem{Jaksch2003}
\bibinfo{author}{D. Jaksch} and \bibinfo{author}{P. Zoller},
\newblock \bibinfo{title}{Creation of effective magnetic fields in optical lattices: the Hofstadter butterfly for cold neutral atoms},
\href{https://iopscience.iop.org/article/10.1088/1367-2630/5/1/356}{
\newblock \bibinfo{journal}{New J. Phys.}
\textbf{\bibinfo{volume}{5}}, \bibinfo{pages}{56}
(\bibinfo{year}{2003})}.


\bibitem{supplementary}
\bibinfo {note} {See Supplemental Material for detailed information on the derivation of multiple-component Hubbard model, the variational Monte Carlo approach, et.al, which includes Refs.~\cite{Imambekov2003, Aidelsburger2011, Jaksch2003, Jotzu2014, Liu2010, Tu2013, Liu2018, Read2000, Liu2015, Cai2009}.}


\bibitem{Liu2010}
\bibinfo{author}{Z.-X. Liu}, \bibinfo{author}{Y. Zhou}, and \bibinfo{author}{T.-K. Ng},
\newblock \bibinfo{title}{Fermionic theory for quantum antiferromagnets with spin $S>\frac{1}{2}$},
\href{https://journals.aps.org/prb/abstract/10.1103/PhysRevB.82.144422}{
\newblock \bibinfo{journal}{Phys. Rev. B}
\textbf{\bibinfo{volume}{82}}, \bibinfo{pages}{144422}
(\bibinfo{year}{2010})}.


\bibitem{Liu2015}
\bibinfo{author}{T. Liu}, \bibinfo{author}{W. Li}, \bibinfo{author}{A. Weichselbaum}, \bibinfo{author}{J. von Delft}, and \bibinfo{author}{G. Su},
\newblock \bibinfo{title}{Simplex valence-bond crystal in the spin-1 kagome Heisenberg antiferromagnet},
\href{https://journals.aps.org/prb/abstract/10.1103/PhysRevB.91.060403}{
\newblock \bibinfo{journal}{Phys. Rev. B}
\textbf{\bibinfo{volume}{91}}, \bibinfo{pages}{060403(R)}
(\bibinfo{year}{2015})}.


\bibitem{Rizzi2005}
\bibinfo{author}{M. Rizzi}, \bibinfo{author}{D. Rossini}, \bibinfo{author}{G. D. Chiara}, \bibinfo{author}{S. Montangero}, and \bibinfo{author}{R. Fazio},
\newblock \bibinfo{title}{Phase Diagram of Spin-1 Bosons on One-Dimensional Lattices},
\href{https://journals.aps.org/prl/abstract/10.1103/PhysRevLett.95.240404}{
\newblock \bibinfo{journal}{Phys. Rev. Lett.}
\textbf{\bibinfo{volume}{95}}, \bibinfo{pages}{240404}
(\bibinfo{year}{2005})}.


\bibitem{Chin2010}
\bibinfo{author}{C. Chin}, \bibinfo{author}{R. Grimm}, \bibinfo{author}{P. Julienne}, and \bibinfo{author}{E. Tiesinga},
\newblock \bibinfo{title}{Feshbach resonances in ultracold gases},
\href{https://journals.aps.org/rmp/abstract/10.1103/RevModPhys.82.1225}{
\newblock \bibinfo{journal}{Rev. Mod. Phys.}
\textbf{\bibinfo{volume}{82}}, \bibinfo{pages}{1225}
(\bibinfo{year}{2010})}.


\bibitem{Papoular2010}
\bibinfo{author}{D. J. Papoular}, \bibinfo{author}{G. V. Shlyapnikov}, and \bibinfo{author}{J. Dalibard},
\newblock \bibinfo{title}{Microwave-induced Fano-Feshbach resonances},
\href{https://journals.aps.org/pra/abstract/10.1103/PhysRevA.81.041603}{
\newblock \bibinfo{journal}{Phys. Rev. A}
\textbf{\bibinfo{volume}{81}}, \bibinfo{pages}{041603(R)}
(\bibinfo{year}{2010})}.


\bibitem{Cai2009}
\bibinfo{author}{Z. Cai}, \bibinfo{author}{S. Chen}, and \bibinfo{author}{Y.~P. Wang},
\newblock \bibinfo{title}{Spontaneous trimerization in two-dimensional antiferromagnets},
\href{https://iopscience.iop.org/article/10.1088/0953-8984/21/45/456009}{
\newblock \bibinfo{journal}{J. Phys.: Condens. Matter}
\textbf{\bibinfo{volume}{21}}, \bibinfo{pages}{456009}
(\bibinfo{year}{2009})}.


\bibitem{Boll2016}
\bibinfo{author}{M. Boll}, \bibinfo{author}{T. A. Hilker}, \bibinfo{author}{G. Salomon}, \bibinfo{author}{A. Omran}, \bibinfo{author}{J. Nespolo}, \bibinfo{author}{L. Pollet}, \bibinfo{author}{I. Bloch}, and \bibinfo{author}{C. Gross},
\newblock \bibinfo{title}{Spin- and density-resolved microscopy of antiferromagnetic correlations in Fermi-Hubbard chains},
\href{https://www.science.org/doi/10.1126/science.aag1635}{\newblock \bibinfo{journal}{Science}
\textbf{\bibinfo{volume}{353}}, \bibinfo{pages}{1257}
(\bibinfo{year}{2016})}.


\bibitem{Kitaev2006L}
\bibinfo{author}{A. Kitaev} and \bibinfo{author}{J. Preskill},
\newblock \bibinfo{title}{Topological Entanglement Entropy},
\href{https://journals.aps.org/prl/abstract/10.1103/PhysRevLett.96.110404}{
\newblock \bibinfo{journal}{Phys. Rev. Lett.}
\textbf{\bibinfo{volume}{96}}, \bibinfo{pages}{110404}
(\bibinfo{year}{2006})}.


\bibitem{Zhang2011L}
\bibinfo{author}{Y. Zhang}, \bibinfo{author}{T. Grover}, and \bibinfo{author}{A. Vishwanath},
\newblock \bibinfo{title}{Entanglement Entropy of Critical Spin Liquids},
\href{https://journals.aps.org/prl/abstract/10.1103/PhysRevLett.107.067202}{
\newblock \bibinfo{journal}{Phys. Rev. Lett.}
\textbf{\bibinfo{volume}{107}}, \bibinfo{pages}{067202}
(\bibinfo{year}{2011})}.


\bibitem{Islam2015}
\bibinfo{author}{R. Islam}, \bibinfo{author}{R. Ma}, \bibinfo{author}{P. M. Preiss}, \bibinfo{author}{M. E. Tai}, \bibinfo{author}{A. Lukin}, \bibinfo{author}{M. Rispoli}, and \bibinfo{author}{M. Greiner},
\newblock \bibinfo{title}{Measuring entanglement entropy in a quantum many-body system}, \href{https://www.nature.com/articles/nature15750}{\newblock \bibinfo{journal}{Nature (London)}
\textbf{\bibinfo{volume}{528}}, \bibinfo{pages}{77}
(\bibinfo{year}{2015})}.


\bibitem{Yan2022}
\bibinfo{author}{Z. Z. Yan}, \bibinfo{author}{B. M. Spar}, \bibinfo{author}{M. L. Prichard}, \bibinfo{author}{S. Chi}, \bibinfo{author}{H.-T. Wei}, \bibinfo{author}{E. Ibarra-Garc\'{\i}a-Padilla}, \bibinfo{author}{K. R. A. Hazzard}, and \bibinfo{author}{W. S. Bakr},
\newblock \bibinfo{title}{Two-Dimensional Programmable Tweezer Arrays of Fermions},
\href{https://journals.aps.org/prl/abstract/10.1103/PhysRevLett.129.123201}{
\newblock \bibinfo{journal}{Phys. Rev. Lett.}
\textbf{\bibinfo{volume}{129}}, \bibinfo{pages}{123201}
(\bibinfo{year}{2022})}.


\bibitem{Kaufman2021}
\bibinfo{author}{A. M. Kaufman} and \bibinfo{author}{K.-K. Ni},
\newblock \bibinfo{title}{Quantum science with optical tweezer arrays of ultracold atoms and molecules},
\href{https://www.nature.com/articles/s41567-021-01357-2}{
\newblock \bibinfo{journal}{Nat. Phys.}
\textbf{\bibinfo{volume}{17}}, \bibinfo{pages}{1324}
(\bibinfo{year}{2021})}.


\bibitem{Tu2013}
\bibinfo{author}{H.-H. Tu},
\newblock \bibinfo{title}{Projected BCS states and spin Hamiltonians for the SO($n$)$_1$ Wess-Zumino-Witten model},
\href{https://journals.aps.org/prb/abstract/10.1103/PhysRevB.87.041103}{
\newblock \bibinfo{journal}{Phys. Rev. B}
\textbf{\bibinfo{volume}{87}}, \bibinfo{pages}{041103}
(\bibinfo{year}{2013})}.


\bibitem{Read2000}
\bibinfo{author}{N. Read} and \bibinfo{author}{D. Green},
\newblock \bibinfo{title}{Paired states of fermions in two dimensions with breaking of parity and time-reversal symmetries and the fractional quantum Hall effect},
\href{https://journals.aps.org/prb/abstract/10.1103/PhysRevB.61.10267}{
\newblock \bibinfo{journal}{Phys. Rev. B}
\textbf{\bibinfo{volume}{61}}, \bibinfo{pages}{10267}
(\bibinfo{year}{2000})}.

\end{thebibliography}
\end{document}